\documentclass[aps,prd,twocolumn,superscriptaddress,floatfix]{revtex4}
\usepackage{amsmath, amsfonts}
\input epsf
\usepackage{graphics}
\usepackage[sort&compress]{natbib}

\begin{document}
\title{Relativistic hydrodynamics in the presence of puncture black holes}
\date{\today}
\author{Joshua A. Faber}
\altaffiliation{National Science Foundation (NSF) Astronomy and
  Astrophysics Postdoctoral Fellow.}
\email{jfaber@uiuc.edu}
\affiliation{Department of Physics, University of Illinois at
  Urbana-Champaign, Urbana, IL 61801}
\author{Thomas W. Baumgarte}
\altaffiliation{Also at Department of Physics, University of Illinois at
  Urbana-Champaign, Urbana, IL 61801}
\affiliation{Department of Physics and Astronomy, Bowdoin College,
  Brunswick, ME 04011}
\author{Zachariah B. Etienne}
\affiliation{Department of Physics, University of Illinois at
  Urbana-Champaign, Urbana, IL 61801}
\author{Stuart~L.~Shapiro}
\altaffiliation{Also at Department of Astronomy and NCSA, University of
  Illinois at Urbana-Champaign, Urbana, IL 61801}
\affiliation{Department of Physics, University of Illinois at
  Urbana-Champaign, Urbana, IL 61801}
\author{Keisuke Taniguchi}
\affiliation{Department of Physics, University of Illinois at
  Urbana-Champaign, Urbana, IL 61801}

\begin{abstract}
Many of the recent numerical simulations of binary black holes in
vacuum adopt the moving puncture approach.  This successful approach
avoids the need to impose numerical excision of the black hole
interior and is easy to implement.  Here we wish to explore how well
the same approach can be applied to moving black hole punctures in the
presence of relativistic hydrodynamic matter.  First, we evolve single
black hole punctures in vacuum to calibrate our BSSN
(Baumgarte-Shapiro-Shibata-Nakamura) implementation 
and to confirm that the numerical solution for the exterior spacetime
is invariant to any ``junk'' (i.e., constraint-violating) initial data
employed in the black hole interior.  Then we focus on relativistic
Bondi accretion onto a moving puncture Schwarzschild black hole as a
numerical testbed for our high-resolution shock-capturing relativistic hydrodynamics scheme.  We
find that the hydrodynamical equations can be evolved successfully in
the interior without imposing numerical excision.  These results help
motivate the adoption of the moving puncture approach to treat the
binary black hole-neutron star problem using conformal thin-sandwich
initial data. 
\end{abstract}
\maketitle

\section{Introduction}

The evolution of spacetimes containing black holes (BHs) has been a
long-standing focus of numerical relativity.  Among other
astrophysical systems containing compact objects, BHs in
binaries with compact object companions have long been considered
promising sources of gravitational waves that can be detected by the
new generation of ground-based gravitational wave interferometers LIGO
\cite{LIGO}, TAMA \cite{TAMA}, GEO \cite{GEO} and VIRGO \cite{VIRGO},
and by the proposed space-based detector LISA \cite{LISA}.

The recent association of short gamma-ray bursts with galaxies with
extremely low star-formation rates has virtually ruled out supernovae
as the progenitors of short gamma-ray bursts, and instead favors
either stellar-mass black hole-neutron
star (BHNS) or neutron star-neutron star (NSNS) mergers (see
\cite{SGRB} for a review).  While multiple aspects of the NSNS
scenario have been studied in detail (see, e.g.,
\cite{STU1,STU2,MDSB,ST,HMNS1,HMNS2,ShiUIUC} for simulations performed
in general relativity), progress on BHNS mergers has been
significantly slower.  This is not surprising, since simulations
of BHNS binaries combine the computational difficulties associated with the
BH singularity with those arising from shocks and other
hydrodynamic phenomena associated with the neutron star matter.

Recent advances in the numerical simulation of BHBH binaries 
(see \cite{FP1,UTB1,God1} as well as numerous
follow-up papers) have
overcome many of the difficulties associated with the BH singularity.
In particular, the ``moving puncture'' approach adopted by
\cite{UTB1,God1} does not require any excision of the BH
interior, is quite easy to implement into BSSN numerical relativity
schemes \cite{ShiN95,BauS99},
and has been used successfully in a large number of simulations.
These simulations have treated the case of equal-mass,
non-spinning binaries \cite{UTB2,God2,God3,God4,Jena3}, and in some cases
have also dealt with
binaries with unequal mass \cite{Godq1,PSU1,Jena1}, non-zero spins either
aligned with the orbital angular momentum
\cite{UTBSpin1,UTBSpin2,PSU2,Marro1} or at arbitrary orientations
\cite{UTB2spins1,Jena2,TM}, or combinations of both
\cite{UTB2spins2,Godq2} (see also
\cite{FP2,FP3,CC1,CC2} for alternative approaches to the BHBH problem).   
These puncture simulations have furnished results of 
great astrophysical interest, including the gravitational wave signals
expected from such mergers and the kick velocities imposed on the merger
remnant. 

In the moving puncture approach the presence of the singularity is
largely ignored, which raises the question as to why it does not spoil the
numerical evolution.  This issue has been clarified by
\cite{JenaGeo1,JenaGeo2}, who analyzed the geometric structure of
puncture solutions.  Under the gauge conditions that are used in
moving puncture simulations, the dynamical evolution of a single
isolated Schwarzschild BH settles down to a time slicing that terminates at a
limiting surface of finite areal radius.  This slicing therefore
avoids the central curvature singularity, and covers only regular
regions of the Schwarzschild geometry.  In effect, the moving puncture
approach provides a means of ``excision-without-excision'' (see also
\cite{Brown,Brown2} and Section \ref{sec:movpuncevolve} below).  An analytic
expression describing the asymptotic, late-time solution has been found in
\cite{BaumPunc}, and we use this solution to test the convergence
behavior of our code in the presence of a puncture BH.  

The fact that time slices in the moving puncture approach cover only
regular regions of spacetime makes it an attractive approach for
modeling BHNS binaries, since then the equations of relativistic
hydrodynamics -- or any other matter model -- can be integrated
together with the gravitational field equations without any need for
excision.  In fact, this is the approach adopted in the only fully
self-consistent dynamical simulations of orbiting BHNS binaries to
date \cite{SUpunc1,SUpunc2} (see \cite{FBST,FBSTR,AEI,NUPSU}
 for other preliminary relativistic simulations of BHNS binaries, and
 \cite{BR,BHR} for discussion about and simulations of alternate
 configurations without the use of excision).

In this paper we analyze in detail the hypothesis that relativistic
hydrodynamics in the presence of puncture BHs does not require
excision.  As a test problem we study relativistic
Bondi accretion onto a BH and compare with the analytic
solution for stationary, spherical flow.  We perform this test both for static and moving black
holes (the latter at rest with respect to the asymptotic gas flow), 
and verify that the fluid behaves as expected.

This paper forms a natural stepping-stone in our group's systematic
effort to model BHNS binaries.  After solving the initial value
problem in general relativity \cite{BauSS04,TanBFS05,TBFS1,TBFS2} and performing
preliminary dynamical simulations in conformal gravitation \cite{FBSTR,FBST}, we
have now assembled and tested the tools to model BHNS binaries
self-consistently.  In contrast to \cite{SUpunc1,SUpunc2}, who
constructed BHNS initial data with a puncture method (see Section
\ref{sec:numerical.ID} below), we plan to evolve our quasiequilibrium
initial data, which we have constructed via the conformal
thin-sandwich method.  

Our forthcoming evolution of conformal thin-sandwich initial data 
raises another conceptional issue that we address in this paper.
To construct quasiequilibrium BH initial data requires excising the
interior and imposing suitable boundary conditions on the BH
horizon \cite{CPInit,Caudill}.  This approach provides data only in the black
hole exterior, but the moving puncture approach for the dynamical
evolution requires data in the BH interior as well.  However,
by definition of the BH horizon, data in the BH
interior cannot affect the exterior spacetime.  We demonstrate in
Section \ref{sec:baumpunc} that we can indeed replace data in
the BH interior with constraint-violating ``junk'' data, and still find
an identical evolution in the exterior.  In fact, the evolution
settles down to the same asymptotic late-time solution in the BH
interior as well, independently of the initial data.  These findings thus
suggest that moving puncture simulations of BHBH binaries can adopt
conformal thin-sandwich initial data, which may be less inherently
eccentric and thus be more accurate approximations to true quasi-equilibrium than the
puncture initial data typically used in moving puncture binary
simulations \cite{BIW}.  This hypothesis is tested in detail in a
forthcoming work \cite{EFLSB}.

This paper is organized as follows.  In Sec.~\ref{sec:numerical}, we
summarize the puncture BH formalism, as well as our specific
implementation of initial data, gauge conditions, matter evolution,
and code diagnostics.  In
Sec.~\ref{sec:vacuum}, we discuss results for a number of vacuum
spacetime evolutions performed for both stationary and moving puncture
BHs, including calculations for which we alter the initial data in the BH
interior.  In Sec.~\ref{sec:matter}, we present simulations of
puncture BH spacetimes that contain hydrodynamic matter, and discuss how we
implement a scheme that allows for stable and accurate
hydrodynamical evolutions.  We conclude in Sec.~\ref{sec:discussion} with a
brief discussion of our findings.

\section{Numerical Implementation}\label{sec:numerical}

\subsection{The 3+1 decomposition}

Throughout this paper we will cast the spacetime metric $g_{ab}$ into
the 3+1 ADM (Arnowitt-Deser-Misner) form
\begin{equation}
ds^2=-(\alpha^2-\beta_i\beta^i)dt^2+2\beta_i dt dx^i +
\psi^4\tilde{\gamma}_{ij}dx^idx^j,
\end{equation}
where $\alpha$ is the lapse function, $\beta^i$ the shift vector,
$\psi$ the conformal factor, and $\tilde{\gamma}_{ij}\equiv \psi^{-4}\gamma_{ij}$ the conformally
related spatial metric, defined so that its determinant
$\tilde{\gamma}=1$ in Cartesian coordinates.  The extrinsic curvature
$K_{ij}$ is defined by
\begin{equation}
(\partial_t-{\cal L}_\beta)\gamma_{ij}=-2\alpha K_{ij},
\end{equation}
where ${\cal L}$ denotes the Lie derivative.  We note that two
independent conformal rescalings of the tracefree part of the
extrinsic curvature $A_{ij}$ are widely used in the literature.  In
the context of initial data decompositions, this quantity is usually
rescaled as
\begin{equation}
\bar{A}_{ij}\equiv \psi^2 \left(K_{ij}-
\frac{1}{3}g_{ij}K\right)\label{eq:aijbar},
\end{equation}
whereas BSSN-style evolution schemes typically rescale $A_{ij}$ as
\begin{equation}
\tilde{A}_{ij}\equiv \psi^{-4} \left(K_{ij}-\frac{1}{3}g_{ij}K\right),
\end{equation}
so that its indices may be raised and lowered with the conformally
related spatial metric.  Our code evolves the latter expression.

We adopt the BSSN formulation \cite{ShiN95,BauS99} for the dynamical
evolution of the gravitational fields (see also Eqs.~(11) -- (15) of
\cite{DMSB}).  In addition to the quantities $\tilde{\gamma}_{ij}$,
$\phi \equiv \ln \psi$, $\tilde{A}_{ij}$ and $K$, this formulation
utilizes the ``conformal connection functions''
$\tilde{\Gamma}^i\equiv -{\tilde{\gamma}^{ij}}_{,j}$ as auxiliary
quantities.  Like many BSSN implementations, we enforce the vanishing
trace of $\tilde{A}_{ij}$ and unit determinant of
$\tilde{\gamma}_{ij}$ at every timestep.

\subsection{Puncture initial data}
\label{sec:numerical.ID}

The basic idea of puncture initial data is to factor out from the
spatial metric analytic terms that represent the singular terms at a
BH singularity, and treat only the remaining regular terms
numerically \cite{BeiO94,BeiO96,BB97}.  Specifically, under the
assumption of conformal flatness and maximal slicing, the momentum
constraint decouples from the Hamiltonian constraint in the conformal
transverse-traceless decomposition of Einstein's constraint equations
(see, e.g., \cite{Coo00,BauS03} for reviews).  Moreover, the momentum
constraint becomes linear and a solution representing a BH
with momentum $P^i$ is given by 
\begin{equation}
\bar{A}^{ij}= \frac{3}{2r^2}[P^i
  n^j + P^j n^i - (\gamma^{ij}-n^i n^j)P^k n_k],\label{eq:aijpunc}
\end{equation}
where $r$ and $n^i$ are the distance to and radial unit normal vector
from the BH, and the expression is scaled as in
Eq.~(\ref{eq:aijbar}).  This expression can then be inserted into the
Hamiltonian constraint.  Decomposing the conformal factor as a sum of an
analytic, singular part $\psi_s\equiv 1+{\mathcal M}/2r$ and a term $u$ that is
regular everywhere,
\begin{equation} \label{eq:confinit}
\psi = u + \psi_s \equiv 1+u + \frac{{\mathcal M}}{2r},
\end{equation}
where ${\mathcal M}$ is a constant,
the Hamiltonian constraint becomes a regular equation for $u$,
\begin{equation}
\nabla^2 u +\frac{1}{8}\psi_s^{-7}
\bar{A}^{ij}\bar{A}_{ij}(1+u/\psi_s)^{-7}=0.\label{eq:boostpunc}
\end{equation}
For moving BHs, the solution for
$\psi$ can be found analytically up to second order in $P^i$ (see,
e.g., \cite{Laguna})
\begin{subequations}
\begin{eqnarray}
u&= &\frac{P^2}{8{\mathcal M}({\mathcal M}+2r)^5}[u_0(r)P_0(\mu)+u_2(r)P_2(\mu)]\label{eq:upunc}\\
u_0&=&{\mathcal M}^4+10{\mathcal M}^3r+40{\mathcal M}^2r^2+80{\mathcal M}r^3+80r^4\label{eq:upunc0}\\
u_2&=&\frac{{\mathcal M}}{5r^3} \Big( 42{\mathcal M}^5r+378{\mathcal M}^4r^2+1316{\mathcal M}^3r^3\nonumber\\
&&+2156{\mathcal M}^2r^4+1536{\mathcal M}r^5+240r^6\nonumber\\
&& +21{\mathcal M}({\mathcal M}+2r)^5\ln\big(\frac{{\mathcal M}}{{\mathcal M}+2r}\big)
\Big) ,\label{eq:upunc2}
\end{eqnarray}
\end{subequations}
where $P_0(\mu)=1$ and $P_2(\mu)=3\mu^2/2-1/2$ are Legendre
polynomials and $\mu\equiv\cos\theta$.  Clearly, for static BHs with
$P^i = 0$ we find $u = 0$ as expected, and recover a Schwarzschild
$t={\rm const.}$ slice in isotropic coordinates.  We also note that
the parameter ${\mathcal M}$ in the above equations reduces to the
BH ADM mass only for static BHs (see also Section
\ref{sec:diagnostics} below).  

\subsection{Moving puncture evolutions}
\label{sec:movpuncevolve}

The success of the puncture method for initial data suggests that a
similar approach might be equally successful for dynamical
evolutions.  However, the original puncture method -- namely, factoring
out analytical singular terms and evolving the remaining regular
terms alone -- did not achieve long-term stable evolutions in dynamical
simulations (see, e.g., \cite{Bru99,AlcBBLNST01}).  The problem may be
associated with the need for a coordinate system that leaves the
puncture at a prescribed location in the numerical grid, given by the
singularity in the analytical function.  The breakthrough in the
recent dynamical puncture simulations is based on using a
``moving'' puncture in which no singular term is factored out
\cite{UTB1,God1}.  With suitable coordinate conditions (see
Section \ref{sec:gauge} below), this prescription
leads to remarkably stable evolutions.  This is somewhat surprising,
since one might have expected the presence of singularities to spoil
the numerical evolution.  This issue has been clarified recently by
\cite{JenaGeo1,JenaGeo2}, who analyzed the geometric structure of
puncture solutions.

For a single isolated BH at rest in the coordinate system, initial data representing a slice of constant
Schwarzschild time in isotropic coordinates are given by
Eq.~(\ref{eq:confinit}) with $u = 0$.  Isotropic coordinates do not penetrate
the BH horizon, and instead cover two copies of the BH
exterior that can be thought of as two different asymptotically flat
universes connected by an Einstein-Rosen bridge.  The conformal factor
Eq.~(\ref{eq:confinit}) diverges at $r = 0$; this point
corresponds to the asymptotically flat end of the ``other'' universe,
and hence is a relatively harmless coordinate singularity only.
This initial slice does not encounter the BH spacetime singularity.

With the gauge conditions typically used in moving puncture
simulations, such initial data lead to an episode of ``dynamical''
evolution until the solutions settle down to a new equilibrium
state.  Specifically, in the adopted gauge the metric coefficients 
depend on time, and hence appear dynamical until they settle down.  In the new equilibrium
state the conformal factor is still singular at $r=0$, but it now
features a $1/\sqrt{r}$ singularity instead of the previous $1/r$ coordinate
singularity.  Such behavior indicates that the point $r=0$ now represents a
limiting surface of finite areal radius $R_{\rm areal}$.  The slice
has thus disconnected from the other asymptotically flat end, and
instead terminates on a surface of finite $R_{\rm areal}$ inside the
BH's horizon.  The numerical grid does not include the
spacetime singularity at $R_{\rm areal} = 0$, so that the point $r=0$
is again a coordinate singularity only.  This helps explain the
success of this numerical method, which, in short, may be thought of
as ``excision-without-excision'' (see also \cite{Brown} for an
explanation of this numerical behavior).  

This realization about puncture geometry suggests that it should be
straightforward to extend the puncture approach to describe
relativistic hydrodynamics in the presence of BHs.  The numerical grid covers
only regular regions of the spacetime, so the fluid cannot encounter
any spacetime singularities.  At $r=0$, which corresponds to a sphere
of finite areal radius $R_{\rm areal}$ inside the horizon, all field
and fluid characteristics point inward to smaller areal radii.  The interior of
the sphere may therefore be disregarded -- it cannot affect the
exterior.  As we will
discuss in Section \ref{sec:hydro} below, finite differencing around
the point $r=0$ can lead to numerical error that complicates some of our
expectations; however, this is a purely numerical issue that can be dealt
with quite straightforwardly.

\subsection{Gauge and boundary conditions}
\label{sec:gauge}
The key to the success of a moving puncture evolution is the
identification of suitable gauge conditions.  Both \cite{UTB1} and 
\cite{God1} use an advective ``1+log'' slicing condition for the lapse
\begin{equation}
\partial_t\alpha-\beta^i\partial_i\alpha=2\alpha K.\label{eq:lapseevol}
\end{equation}
The advection term forces the singularity, located at the point where
$\psi\rightarrow\infty$ and $\alpha\rightarrow 0$, to move with
coordinate velocity $v^i\equiv dx^i/dt=-\beta^i$.  We note that we
evaluate all advective terms $\beta^i\partial_i$ using ``upwind''
differencing, as described in \cite{ShiUpwind}.

The gauge condition used in moving puncture evolutions is a
hyperbolic ``gamma-driver'' condition for the shift, either
\begin{subequations}
\begin{eqnarray}
\partial_t\beta^i -\beta^j\partial_j \beta^i &=& \frac{3}{4} B^i,
\label{eq:sshiftevol1}\\
\partial_tB^i -\beta^j\partial_j B^i &=& \partial_t
\tilde{\Gamma}^i -\beta^j\partial_j \tilde{\Gamma}^i-\eta B^i,
\label{eq:sshiftevol2}
\end{eqnarray}
\end{subequations}
or
\begin{subequations}
\begin{eqnarray}
\partial_t\beta^i &=& \frac{3}{4} B^i,\label{eq:shiftevol1}\\
\partial_tB^i &=& \partial_t \tilde{\Gamma}^i-\eta B^i.\label{eq:shiftevol2}
\end{eqnarray}
\end{subequations}
The former choice is referred to in \cite{GodGauge} as the
``shifting-shift'' condition, and represents the choice
$\zeta_S=\zeta_b=\zeta_f=0$ in the notation of \cite{Gundlach}, who
studied the hyperbolicity of the resulting evolution scheme.  The
latter is referred to as the ``non-shifting-shift'', and has
$\zeta_S=\zeta_b=\zeta_f=1$.  We note that \cite{SUpunc1} use a
modified, first-order version of the non-shifting-shift with a
different stabilization term.  Throughout this work, we will use only
the ``non-shifting-shift'', Eqs.~(\ref{eq:shiftevol1}) and
(\ref{eq:shiftevol2}).

The term $\eta$ in Eqs.~(\ref{eq:sshiftevol2}) and
(\ref{eq:shiftevol2}), which has units of ${\mathcal M}^{-1}$, is a
damping term that has a non-trivial effect on the evolution of the
puncture as it moves across the grid.  In general, for larger values
of $\eta$, the BH has a smaller coordinate velocity but a larger
coordinate radius, which allows for better numerical resolution of the
BH horizon. However, larger values of $\eta$ were shown in
\cite{GodGauge} to produce significantly larger values of
$\tilde{\Gamma}^i$, indicating a stronger deformation of the metric
during the evolution.  This is often disadvantageous, especially with
regard to specifying boundary conditions.  We use $\eta=0.25/{\mathcal
M}$ for the static BHs in Sec.~\ref{sec:baumpunc},
$\eta=0.2/{\mathcal M}$ for matter evolutions in
Sec.~\ref{sec:matter}, but $\eta=2.0/{\mathcal M}$ for moving black
holes in Secs.~\ref{sec:statbh} and \ref{sec:movpunc} so that we can
directly compare with the results of \cite{GodGauge}.

For our runs with relatively close outer boundaries described in
Sec.~\ref{sec:baumpunc} we use Robin-type boundary conditions for our
metric and field variables, assuming a $1/r$ falloff behavior away
from their asymptotic values.  For larger runs we employ outgoing
wavelike boundary conditions, based on the local speed of light and
again assuming a $1/r$ falloff.  The conditions are slightly modified
when we use a ``fisheye'' coordinate scheme, as we discuss in
Appendix~\ref{app:fisheye}.

\subsection{Hydrodynamics}
\label{sec:hydro}

Our hydrodynamics scheme is essentially identical to the grid-based, fully
general relativistic, high resolution shock capturing scheme
described in \cite{DLSS}, although here we do not include any
electromagnetic terms in our evolution calculations. This scheme for
both matter and field evolution is
based on second-order finite differencing, which we have now embedded
within the {\tt Cactus} grid hierarchy \cite{Cactus}.

We assume a
stress energy tensor in the form $T^{\mu\nu}=\rho_0 h u^\mu
u^\nu+Pg^{\mu\nu}$ where $\rho_0$ is the rest density, $h$ the
specific enthalpy, and $u^\mu$ the 4-velocity of the fluid.  We will
also assume a gamma-law equation of state
\begin{equation}
P=(\Gamma-1)\rho_0\epsilon,
\end{equation}
where $\epsilon$ is the specific internal energy.
 
In our numerical code, we evolve the ``conserved'' quantities
$\rho_*$, $\tilde{S}_i$, and $\tilde{\tau}$, defined by Eqs.~(36) and
(37) of \cite{DLSS}.  These conserved quantities are defined in terms
of the ``primitive'' variables $\rho$, $v^i$, and $P$, according to
their Eqs.~(41) -- (43).  The transformation from the conserved
quantities back to the primitive set requires an iteration (see
Eqs.~(59) -- (62) of \cite{DLSS}).  

We find that this iteration is very accurate everywhere except in the
region immediately surrounding the puncture, where accumulated errors
in the finite differencing can lead to unphysical values of the
primitive variables.  
To overcome this difficulty, we note that the quantity $\tilde{\tau}$, which serves as the
energy variable in the conserved set of variables, should always
remain positive. In fact, it has to remain greater than $w-\rho_*$,
where $w\equiv \alpha u^0\rho_*$, in order to satisfy $h\ge 1$ (see
Eq.~(58) in \cite{DLSS}) , but positivity is sufficient for our purposes.  To 
enforce this, we set $\tilde{\tau}$ to a minimum value, typically 
$\tilde{\tau}=10^{-3}$ whenever it would otherwise be negative.
Furthermore, from (58) in \cite{DLSS} we see
that in the limit $h\rightarrow 1$ we have $w\rightarrow
\tilde{\tau}+\rho_*$, or equivalently 
\begin{equation} 
|\tilde{S}|^2\equiv\gamma^{ij}\tilde{S}_i\tilde{S}_j \rightarrow 
\tilde{\tau}(\tilde{\tau}+2\rho_*).
\end{equation}
It can be shown that this limit on $|\tilde{S}|^2$ is an upper limit.
When numerical error leads to a value of $|\tilde{S}|^2$ that is
larger than the allowed value, we enforce this condition by reducing
the magnitude of $\tilde{S}_i$ so that $|\tilde{S}|^2$ is reduced to
$0.98\times \tilde{\tau}(\tilde{\tau}+2\rho_*)$, leaving the ratios of
the individual vector components unchanged.

We again point out that the above ``fixes'' are employed only in the
immediate vicinity of the puncture, inside the BH horizon.
The need for these fixes arises from poor resolution around the
puncture and the resulting large finite differencing error.  We find
that increasing the grid resolution decreases the size of the
region in which the fixes are needed.  Other than dealing with this
break-down in the iteration between conserved and primitive fluid
variables, the equations of relativistic hydrodynamics can be
integrated without need for excision in the presence of puncture black
holes.

\subsection{Coordinates}

We perform simulations both in 2+1 (i.e., axisymmetry)
and in 3+1, using the HRSC, BSSN code of \cite{DLSS}, which
accommodates both cases.  Our 2+1 simulations adopt the
"cartoon" method for the fields, as described in \cite{Cartoon}. 

Finite difference simulations with fixed spatial resolution in three
spatial dimensions are very intensive computationally, since they
simultaneously require a fine grid resolution close to the BH, and
boundary conditions imposed at a sufficiently large 
separation.  We have implemented so-called ``fisheye'' coordinates to
solve the problem of dynamic range (see
\cite{UTB1}).  In Appendix~\ref{app:fisheye} we review the coordinate
transformation and its effect on the various terms that appear in our
numerical scheme, especially our shift evolution equation.

\subsection{Diagnostics}
\label{sec:diagnostics}

We monitor several global quantities computed from surface
integrals at large separations.
These include the ADM mass
\begin{equation}
M_{\rm ADM}=\frac{1}{2\pi}\int (\tilde{\Gamma}^i/8 -
\gamma^{ij}\partial_j\psi)d\Sigma_i,\label{eq:madm}
\end{equation}
and the linear momentum
\begin{equation}
P_j=\frac{1}{8\pi}\oint (K^i_j-\delta^i_j K)d\Sigma_i,\label{eq:padm}
\end{equation}
where $d\Sigma_i= (x^i/r) \psi^6 r^2\sin\theta d\theta d\phi$ for a
spherical surface at fixed radius. 

We also compute the irreducible mass of the BH from
\begin{equation}
M_{\rm irr}=\sqrt{{\mathcal A}/16\pi}. \label{eq:mirr}
\end{equation}
Instead of computing the proper area ${\mathcal A}$ of the BH's event
horizon, we approximate this area as that of the apparent horizon.

For the approximate boosted BH solutions of Section
\ref{sec:numerical.ID} the ADM mass and irreducible mass are related
to the parameter ${\mathcal M}$, to leading order in $P^i$, by
\begin{equation} \label{eq:mass_translate}
M_{\rm ADM} = {\mathcal M} + \frac{5}{8} \, \frac{P^2}{\mathcal M},~~~~~~~
M_{\rm irr} = {\mathcal M} + \frac{1}{8} \, \frac{P^2}{\mathcal M},
\end{equation}
(see \cite{DenBP06}).  Throughout this paper we express dimensional
quantities in units of ${\mathcal M}$.

\section{Vacuum Tests}\label{sec:vacuum}

To test the vacuum sector of our puncture code, we have performed a suite of
tests for stationary and moving puncture BHs, both in 2+1 and 3+1.
In these tests we adopt two slightly different slicing
conditions, namely the 1+log condition Eq.~(\ref{eq:lapseevol}), both with
and without the advective term $\beta^i \partial_i \alpha$.  Most
commonly, moving puncture solutions employ condition
Eq.~(\ref{eq:lapseevol}) with the advective term; we will refer to this
condition as ``advective 1+log'' slicing.  The geometry of the
resulting spacetime has been analyzed in \cite{JenaGeo1}.  In
\cite{JenaGeo2}, the authors considered dropping the advective term,
which results in the ``non-advective 1+log'' slicing
\begin{equation}
\partial_t\alpha=-2\alpha K.\label{eq:lapsebaum}
\end{equation}
As it does for the advective 1+log slicing, the dynamical evolution
using the modified slicing condition quickly
settles down to a new equilibrium, but in this case this
time-independent solution must evidently be maximally sliced with $K=
0$.  As it turns out, this maximally sliced asymptotic solution can be
found analytically (see \cite{BaumPunc}).  In Section
\ref{sec:baumpunc} we therefore perform tests comparing with the
analytic solution, before considering the more common advective 1+log
slicing in Section \ref{sec:advect}.

\subsection{Non-advective 1+log slicing}
\label{sec:baumpunc}

As demonstrated in \cite{JenaGeo2}, dynamical puncture evolutions with
the non-advective 1+log slicing condition Eq.~(\ref{eq:lapsebaum})
settle down to a maximally sliced, time-independent solution, given
analytically by Eqs.~(11)-(15) of \cite{BaumPunc}.  We test 
our code by comparing the late-time solution of dynamical puncture
evolutions with the analytical solution.  In these tests we consider
three different types of initial data, which we summarize in Table
\ref{table:zachruns}.  In the first set of runs, labeled by ``BN''
in Table \ref{table:zachruns}, we adopt the analytical solution of
\cite{BaumPunc} itself as initial data.  The second set of runs,
labeled by ``ISO'', adopts as initial data a $t={\rm const.}$
time slice of the Schwarzschild solution in isotropic coordinates, given by
Eq.~(\ref{eq:confinit}) for $u = 0$, together with $\alpha=\psi^{-2}$.
The third set of initial data is identical to the second set in the BH
exterior, but now we replace the interior data
with some constraint-violating ``junk''.  We adopt these
data to test whether we can use initial data that only provide
valid exterior data in puncture evolutions, which also require initial
data to fill the arrays in the
interior.  We consider three different choices for the
interior solution.  In the first choice, labeled ``Horizon Junk'' or
``HJ'', we set the values of the conformal factor and lapse function
throughout the interior of the BH to their values on the horizon,
i.e.~$\psi=2$ and $\alpha=1/4$ for $r < r_h \equiv {\mathcal M}/2$. 
For the second choice, labeled ``Interior Junk'' or ``IJ'', we again
set $\psi$ and $\alpha$ to constant values, but this time only inside
$r_h/2$, so $\psi = 3$ and $\alpha = 1/9$ for $r < r_h/2$, using the
isotropic solution everywhere outside this.  Finally,
in our third choice, labeled ``Smooth Junk'' or ``SJ'', we construct
an even, fourth-order polynomial in the BH interior that
joins on to the exterior conformal factor in such a way that the
function and its first two derivatives are continuous on the horizon.
Specifically, we choose $\psi=2.875-5(r/{\mathcal M})^2+6(r/{\mathcal
M})^4$ in the BH interior, and $\alpha = \psi^{-2}$.

All simulations described in this Section were performed using
equatorial symmetry.  For 3+1 runs we used fisheye coordinates with
parameters $a_0=1$, $a_1=4$, $R_1=3$, $s_1=1.0$.  While we always
employ a cubic coordinate grid, a cubic grid in fisheye coordinates
does not correspond to a cubic grid in physical coordinates.  Thus,
the boundaries we quote in physical coordinates only apply at the
coordinate axes, and lie further out at other angles. When we quote a
numerical resolution for a given run, it refers to the central region
of the fisheye transition, for which the physical and fisheye
coordinates essentially overlap.  The numerical resolution with
respect to physical coordinates will be more coarse at larger radii,
by construction.  In all calculations described in this section, we
set $\eta=0.25/{\mathcal M}$ in the shift evolution equation. All
evolutions described in this Section were terminated at
$t_F=300{\mathcal M}$.

\begin{table}
\caption{Summary of our stationary vacuum puncture simulations discussed in
Sec.~\protect\ref{sec:baumpunc}.  ``Baumgarte \& Naculich'' refers to the analytical
solution of \protect\cite{BaumPunc}; our different ``junk'' initial 
data are described in the text.}
\begin{tabular}{l|l|l|l}
\hline
Name & Initial Data & Resolution & Grid  \\
\hline 
\multicolumn{4}{c}{2+1; no fisheye; $|x^i|\le 15{\mathcal M}$}\\
\hline
BN2a & Baumgarte \& Naculich & ${\mathcal M}$/32 & $480\times 480$  \\
BN2b & Baumgarte \& Naculich & ${\mathcal M}$/24 & $360\times 360$  \\
BN2c & Baumgarte \& Naculich & ${\mathcal M}$/16 & $240\times 240$  \\
BN2d & Baumgarte \& Naculich & ${\mathcal M}$/8 & $120\times 120$  \\
HJ2a & Isotropic ``Horizon Junk''      & ${\mathcal M}$/32 & $480\times 480$  \\
ISO2a & Isotropic Schwarzschild & ${\mathcal M}$/32 & $480\times 480$  \\
\hline
\multicolumn{4}{c}{3+1; fisheye, $|\bar{x}^i|\le
  6{\mathcal M} \rightarrow |x^i|< 15{\mathcal M}$}\\
\hline
ISO3 & Isotropic Schwarzschild & ${\mathcal M}$/16 & $192^2\times 96$  \\
IJ3 & Isotropic ``Interior Junk'' & ${\mathcal M}$/16 & $192^2\times 96$  \\
HJ3 & Isotropic ``Horizon Junk'' & ${\mathcal M}$/16 & $192^2\times 96$  \\
SJ3 & Isotropic ``Smooth Junk'' & ${\mathcal M}$/16 & $192^2\times 96$  \\
\hline
\end{tabular}
\label{table:zachruns}
\end{table} 

\begin{figure}[ht!]
 \centering \leavevmode \epsfxsize=3in \epsfbox{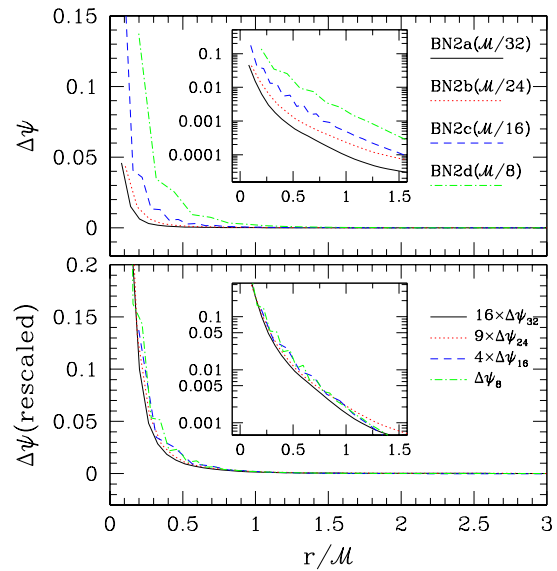}
 \caption{Deviations from the exact solution for the conformal factor
   $\Delta\psi$ for the runs BN2 (see Table \ref{table:zachruns}).
   The top panel shows the deviations themselves, while in the lower
   panel the deviations are appropriately rescaled to demonstrate
   second-order convergence.} 
\label{fig:baum_converge}
\end{figure}

We first compare the ``BN2'' simulations at four different grid
resolutions to establish second-order convergence of our code.  In the
top panel of Fig.~\ref{fig:baum_converge}, we show the absolute
deviations between our numerical runs at $t=10{\mathcal M}$ 
and the exact solution for run BN2a at resolution
${\mathcal M}/32$ (solid), BN2b at ${\mathcal M}/24$ (dotted), BN2c at
${\mathcal M}/16$ (dashed), and BN2d at ${\mathcal M}/8$ (dot-dashed),
focusing on the inner region of the grid, where deviations are most
apparent, and which, at $t = 10 {\mathcal M}$, is causally disconnected
from the outer boundary at $15 {\mathcal M}$.  To confirm the expected
second-order convergence we show the same quantities in the bottom
panel rescaled by factors of 16, 9, 4, and 1, respectively.

\begin{figure}[ht!]
 \centering \leavevmode \epsfxsize=3in \epsfbox{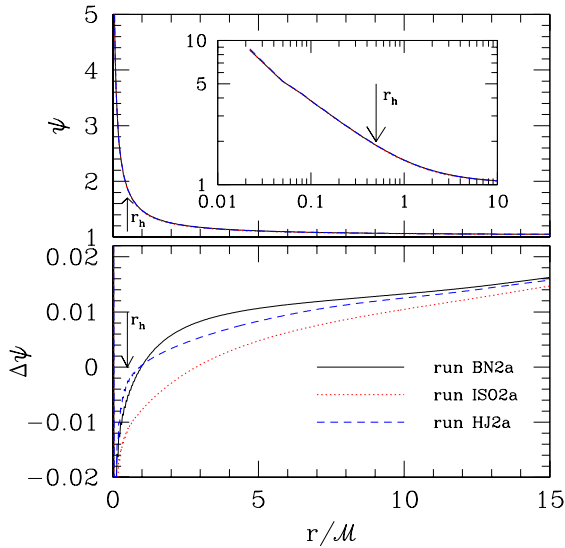}
 \caption{The top panel shows profiles of the conformal factor $\psi$
 at $t=300{\mathcal M}$ for for runs BN2a, ISO2a and HJ2a (see Table
 \ref{table:zachruns}).  All simulations are performed with a grid
 spacing of ${\mathcal M}/32$, using the gauge condition
 Eq.~(\protect\ref{eq:lapsebaum}), for which the late-time solution is
 known exactly.  In the bottom panel we show the absolute deviations
 between our numerical results and the exact solution, finding that
 they are small throughout.  The arrows mark the location of the
 apparent horizon.}
\label{fig:baumpunc}
\end{figure}

We now compare the late-time solution as it emerges from our three
different types of initial data.  In Fig.~\ref{fig:baumpunc} we show
results for runs BN2a, ISO2a and HJ2a, which all have the same grid
resolution.  Evidently, all late-time solutions agree very well with
the analytical solutions.  The deviation from the analytical solution
is at most about 1\% for all three runs, and is caused by finite
difference error as well as the outer boundary.  To within these
numerical errors, all three late-time solutions agree with each other,
even if we replace the interior initial data with junk that does not
even satisfy the constraints.  It is particularly noticeable that even
in the BH interior no ``memory'' of this junk remains at this
late time, and that the solution approaches the analytically known
late-time solution throughout.

\begin{figure}[ht!]
 \centering \leavevmode \epsfxsize=3in \epsfbox{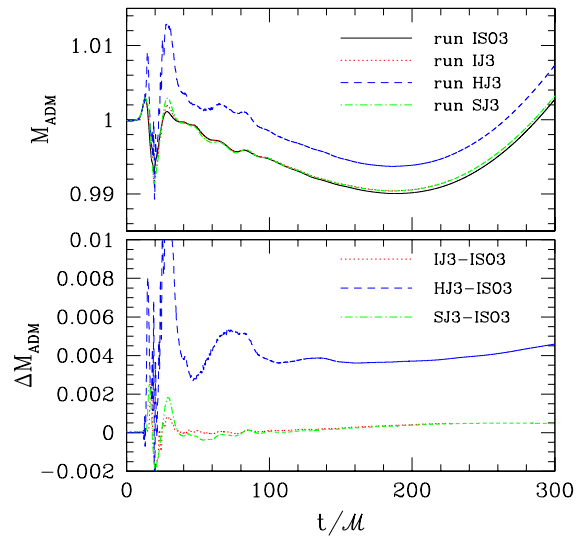}
 \caption{The ADM mass (\protect\ref{eq:madm}) for the 3+1
runs using fisheye and ${\mathcal M}/16$ numerical resolution in the
central region (top panel).  We show results for initial data ISO3,
IJ3, HJ3 and SJ3 (see Table \ref{table:zachruns}).  In the bottom
panel, we plot the differences between the ADM mass from the junk runs
and that of run ISO3.  We see a deviation of $.05\%$ between run ISO3
and runs IJ3 and SJ3 over
the course of the run, indicating that the evolution is insensitive to
the initial data in the BH interior.}
\label{fig:baum_admjunk}
\end{figure}

\begin{figure}[ht!]
 \centering \leavevmode \epsfxsize=3in \epsfbox{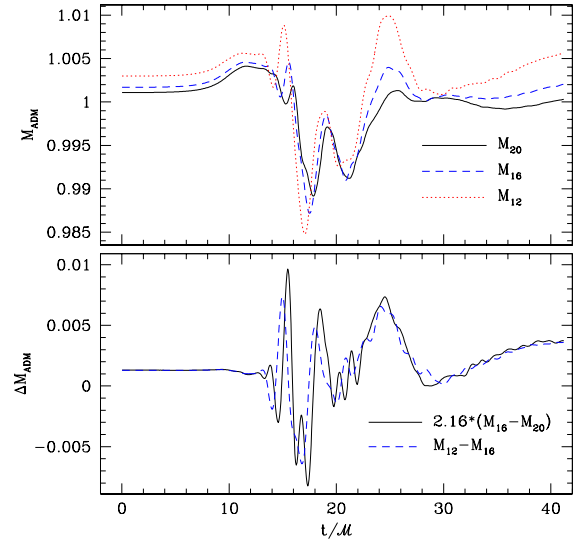}
 \caption{The top panel shows the ADM mass measured at a radius
 $r=12.1{\mathcal M}$, for runs with smooth junk as initial data and 
numerical resolutions of ${\mathcal M}/20$, ${\mathcal M}/16$, and  
${\mathcal M}/12$.  In the bottom panel, the differences between the runs
 are rescaled to demonstrate second-order convergence.}
\label{fig:baum_madmconv}
\end{figure}

While it is not surprising that data in the BH interior does
not affect the exterior solution, it is reassuring to see that this
holds true even in a numerical simulation.  To further illustrate
this point we performed the 3+1 simulations ISO3, HJ3, IJ3, and SJ3,
and plot the ADM mass as a function of time in the top panel
Fig.~\ref{fig:baum_admjunk}.  We find excellent agreement with the
stationary solution at late times for these runs, indicating that
information contained in the BH interior does not affect the
exterior solution as we evolve. In quantitative terms, applying junk
either within the horizon or smoothly at the horizon results in a
change in the ADM mass of $0.05\%$, as we show in the bottom
panel of the Figure. 
We note that reflections off the outer boundary
are at least partially responsible for the discrepancy, given the time
at which they appear.   
To confirm the proper behavior of our code when we have constraint-violating
junk in the BH interior, we show in Fig.~\ref{fig:baum_madmconv} the
convergence of the ADM mass for stationary puncture runs using smooth
junk, performed at numerical resolutions of ${\mathcal M}/20$,
${\mathcal M}/16$, and ${\mathcal M}/12$.  The lower panel suggests self-consistent second-order convergence at
all times.  At both early and late times, we have also verified that the
code converges at second-order to the analytic solution ($M_{\rm ADM}=1$); at
intermediate times we find a brief departure from this behavior that is
likely caused by reflections off the outer boundary.
Even for the simulation HJ3, where the ``junk''
joins the exterior with discontinuous derivatives on the horizon, the
ADM mass deviates from the ISO3 simulations by only $1\%$ over the
course of the evolution.  These simulations indicate that the evolution
is relatively insensitive to the details of the initial data in the BH
interior, even when the differencing stencil in the exterior of the BH
overlaps non-differentiable regions.  Still, our results clearly
suggest that only smoothly extrapolated data should be used in
dynamical evolution calculations.

Our ``junk'' tests suggest a very simple solution to the following
conceptional issue.  Many sets of initial data describing compact
binaries are constructed with the conformal thin-sandwich
decomposition.  For the construction of BHs, these equations
are supplemented with equilibrium boundary conditions that are imposed
on the BHs' horizons.  As a consequence, these data describe
only the exterior geometry, and do not provide any initial data for
the BH interior.  Evolution calculations that employ the
moving puncture approach do not excise the BHs, and hence
require initial data that extend into the BH interior.  Our
tests suggest that we can nevertheless use excised initial data,
e.g., conformal thin-sandwich data, and
simply fill the BH interior with some arbitrary but
sufficiently smooth ``junk''.  This is the approach that we plan to
adopt for simulations of mixed BHNS binaries, enabling us to use our
quasi-equilibrium models \cite{TBFS1,TBFS2} as initial data.

\subsection{Advective 1+log slicing}
\label{sec:advect}

The lapse evolution equation used in Sec.~\ref{sec:baumpunc}
above is useful since it leads to an analytically known solution for a
single BH.  However, most moving puncture simulations of binaries employ
the advective 1+log slicing Eq.~(\ref{eq:lapseevol}) which allows the
punctures to move through the numerical grid.  We now study evolutions
with this advective 1+log slicing, both for stationary and moving
BHs.  Our simulations are summarized in
Table~\ref{table:vacruns}, where ``SP'' refers to stationary puncture
solutions, described in more detail in Section \ref{sec:statbh}, and
``MP'' stands for moving puncture solutions, described in
Sec.~\ref{sec:movpunc}.  For purposes of comparison comparison with
the results of \cite{GodGauge} (see Sec.~\ref{sec:movpunc}) we set
$\eta=2.0/{\mathcal M}$ in Eq.~(\ref{eq:shiftevol2}) for all of the
results discussed in this Section.

\begin{table}
\caption{Vacuum evolution runs discussed in
  Secs.~\protect\ref{sec:statbh} and
  \protect\ref{sec:movpunc}.}
\begin{tabular}{l|l|l|l|l}
\hline
Name & Initial Data & Grid & Resolution & $t_F$ \\
\hline\hline
\multicolumn{5}{c}{2+1 w/Fisheye; $|\bar{x}_{fish}^i|\le 15{\mathcal M}$, $|x^i|\le 136{\mathcal M}$}\\
\hline
SP2a  & Stat.~Punc. & $360\times360$ & ${\mathcal M}$/24 & $100{\mathcal M}$\\
SP2b  & Stat.~Punc. & $240\times240$ & ${\mathcal M}$/16 & $100{\mathcal M}$\\
SP2c  & Stat.~Punc. & $120\times120$ & ${\mathcal M}$/8 & $100{\mathcal M}$\\
\hline
\multicolumn{5}{c}{3+1 w/Fisheye; $|\bar{x}_{fish}^i|\le 12{\mathcal M}$, $|x^i|\le 88{\mathcal M}$}\\
\hline
SP3a & Stat.~Punc. & $240^2\times 120$ & ${\mathcal M}$/10 & $100{\mathcal M}$\\
SP3b & Stat.~Punc. & $192^2\times 96$ & ${\mathcal M}$/8 & $100{\mathcal M}$\\
SP3c & Stat.~Punc. & $144^2\times 72$ & ${\mathcal M}$/6 & $100{\mathcal M}$\\
\hline
\multicolumn{5}{c}{2+1 w/o Fisheye; $P^i=0.5$, $|x^i|\le 70{\mathcal M}$}\\
\hline
MP2a\footnote{For this run, $|x^i|\le 55{\mathcal M}$} & Mov.~Punc. & $1760\times3520$ & ${\mathcal M}$/32 & $50{\mathcal M}$\\
MP2b & Mov.~Punc. & $1680\times3360$ & ${\mathcal M}$/24 & $50{\mathcal M}$\\
MP2c & Mov.~Punc. & $1120\times2240$ & ${\mathcal M}$/16 & $50{\mathcal M}$\\
MP2d & Mov.~Punc. & $560\times1120$ & ${\mathcal M}$/8 & $50{\mathcal M}$\\
\hline
\multicolumn{5}{c}{3+1 w/Fisheye; $P^i=0.5$, $|\bar{x}_{fish}^i|\le 14{\mathcal M}$, $|x^i|\le 120{\mathcal M}$}\\
\hline
MP3a & Mov.~Punc. & $336^2\times 168$ & ${\mathcal M}$/12 & $50{\mathcal M}$\\
MP3b & Mov.~Punc. & $280^2\times 140$ & ${\mathcal M}$/10 & $50{\mathcal M}$\\
MP3c & Mov.~Punc. & $224^2\times 112$ & ${\mathcal M}$/8 & $50{\mathcal M}$\\
MP3d & Mov.~Punc. & $168^2\times 84$ & ${\mathcal M}$/6 & $50{\mathcal M}$\\
\hline
\end{tabular}
\label{table:vacruns}
\end{table}

\subsubsection{Stationary black holes}
\label{sec:statbh}

For our stationary puncture simulations we choose a double fisheye
coordinate system with parameters $a_0=1$, $a_1=4$, $a_2=16$,
$R_1=4.5{\mathcal M}$, $R_2=7.5{\mathcal M}$, and
$s_1=s_2=1.5{\mathcal M}$ (see Appendix \ref{app:fisheye}).  
Our initial data are $t={\rm const.}$
slices of the Schwarzschild metric expressed in isotropic coordinates,
like the ``ISO'' data of Table \ref{table:zachruns}.  We adopt the
gauge conditions Eqs.~(\ref{eq:shiftevol1}) and
(\ref{eq:shiftevolfish}) and evolve to a time $t_F=100{\mathcal M}$.
By this time the solution has settled down into the late-time
equilibrium solution, and further evolution would lead to only very
small changes in the fields.  We use three different grid resolutions
for both our 2+1 and 3+1 simulations to study the convergence behavior
of our code as indicated in Table~\ref{table:vacruns}.  

\begin{figure}[ht!]
 \centering \leavevmode \epsfxsize=3in \epsfbox{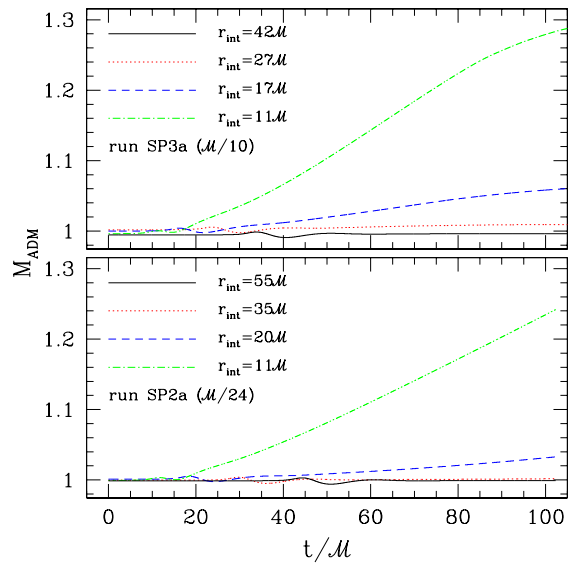}
 \caption{ADM mass versus time for stationary puncture evolutions,
 measured using Eq.~(\protect\ref{eq:madm}) at surfaces of different
 radii.  In the top panel, we show results for surfaces at constant
 fisheye radius $\bar{r}=5.8$ (dot-dashed), $6.8$ (dashed), $7.9$
 (dotted), and $9.0$ (solid) in 3+1 run SP3a, which correspond to
 physical radii $r_{\rm int}=11{\mathcal M}$, $17{\mathcal M}$, $27{\mathcal M}$, and $42{\mathcal M}$,
 respectively.  In the bottom panel, the integration surfaces are
 placed at $\bar{r}=5.8$, $7.2$, $8.5$, and $9.9$, corresponding to
 physical radii $r_{\rm int}=11.1{\mathcal M}$, $19.9{\mathcal M}$, $35.2{\mathcal M}$, and $55.2{\mathcal M}$ in
 2+1 axisymmetric run SP2a.  For the outermost surfaces, we see
 deviations of less than half a percent over the course of the entire
 evolution in each case.}
\label{fig:statpunc_madm}
\end{figure}

In the top panel of Fig.~\ref{fig:statpunc_madm}, we show the ADM mass
for our highest resolution 3+1 calculation, run SP3a, measured at
$\bar{r}_{\rm int}=5.8$, $6.8$, $7.9$, and $9.0$.  Similarly, in the bottom
panel we show the ADM mass for our highest resolution axisymmetric
calculation, run SP2a.
We see two phenomena: first, at a time
roughly corresponding to $t=r_{\rm int}$, we see a small, temporary glitch in
the ADM mass for each value of $r_{\rm int}$, followed by a slow
deviation from the exact value.  This indicates the passage of
junk gravitational radiation, present in the initial data only because
of numerical errors associated with discretization, through the
surface, and is generic to calculations like these.  Second, we note
that our results converge as we move the integration surface outward,
as we would expect.  This is true both before and after the passage of
the junk radiation through the surface.  For the outermost surface, we
see a variation in the ADM mass of no more that half a percent over an
integration time of $100{\mathcal M}$. In general, we find that the
late time deviations from the exact ADM mass scale like $r_{\rm
int}^{-4}$ through most of our grid.

\begin{figure}[ht!]
 \centering \leavevmode \epsfxsize=3in \epsfbox{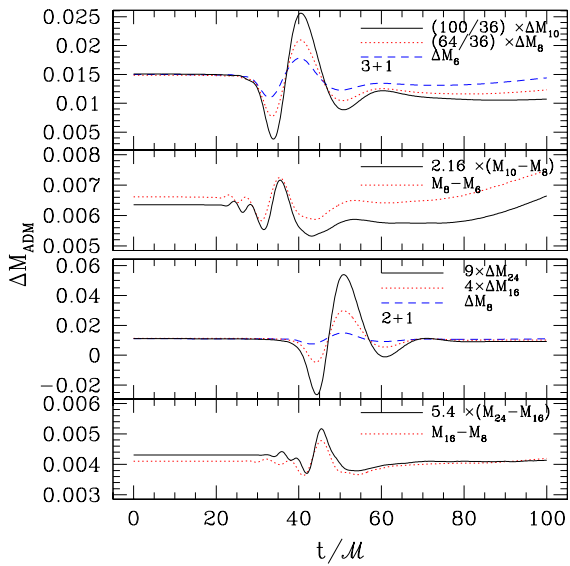}
 \caption{Differences in exact and numerical values for the ADM mass
 for our SP simulations of different numerical
 resolution as a function of time (see Table \ref{table:vacruns}).  In
 the top panel we show results for 3+1 simulations, with the ADM mass
 computed at $r_{\rm int}=41.6 {\mathcal M}$.  In the upper subpanel,
results for different
 resolutions are properly rescaled to demonstrate second-order
 convergence against the exact solution.  In the lower subpanel, we
 compare the differences between pairs of runs to demonstrate
 second-order convergence even during the passage of a spurious
 oscillation caused by initial errors through the data at $r_{\rm int}$ .
The bottom panel shows similar results for 2+1
 simulations, with the ADM mass computed at $r_{int}=55.2 {\mathcal
 M}$.  Conventions are as above.}
\label{fig:statpunc_madmconv}
\end{figure}

To confirm convergence, we calculate the
difference of the ADM mass from its analytical value as a function of
time for all the stationary puncture runs described in
Table~\ref{table:vacruns}.  In the 3+1 case, we take data from the
integration surface placed at $r_{\rm int}=41.6$, whereas for the
2+1 case we choose the surface with $r_{\rm int}=55.2$.  In
Fig.~\ref{fig:statpunc_madmconv}, we see the proper behavior for both
the 3+1 and 2+1 runs (top and bottom panels, respectively), confirming
that our field evolution is indeed convergent to second order in the
grid spacing.  In the upper sub-panels, we show the expected
convergence against the exact solution.  In both the 3+1 and 2+1
simulations, numerical errors inherent in the finite-differenced
initial data reach the integration surfaces at a   
time $t \approx r_{\rm int}$.  During the passage of oscillation, we
still see second-order convergence in the differences between runs,
plotted in the bottom sub-panels of each figure.  
Note that in the top panel, we rescale the higher resolution pair by a factor
$(6^{-2}-8^{-2})/(8^{-2}-10^{-2})=2.16$, whereas in the bottom panel
the rescaling factor is $(8^{-2}-16^{-2})/(16^{-2}-24^{-2})=5.4$.

\begin{figure}[ht!]
 \centering \leavevmode \epsfxsize=3in \epsfbox{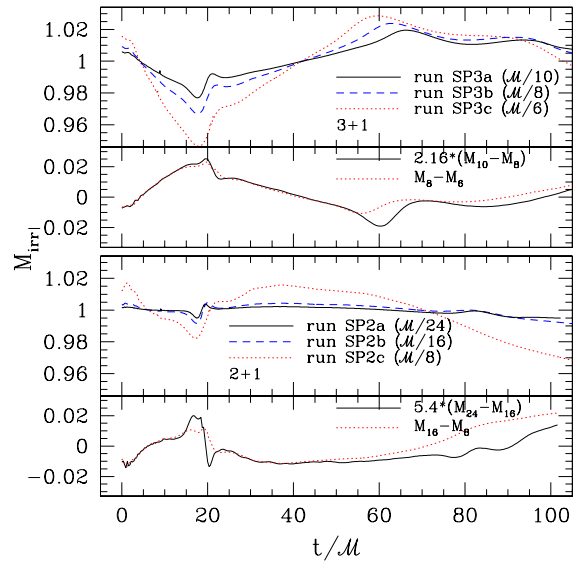}
 \caption{Irreducible mass as a function of time for our 
SP runs (upper subpanels; see Table \ref{table:vacruns}).  For our highest
 resolution runs in 3+1 and 2+1, we see deviations of at most about
 $2\%$ and $0.5\%$, respectively.  In the lower subpanels, we show the
 rescaled differences between pairs of runs, demonstrating
 second-order convergence.}
\label{fig:statpunc_mirr}
\end{figure}

Finally, in Fig.~\ref{fig:statpunc_mirr}, we show the irreducible mass
(see Eq.~(\ref{eq:mirr})) of the BH as a function of time for our 3+1
(top panel) and 2+1 (bottom panel) runs, for all three numerical
resolutions in both cases.  To determine the location and parameters
describing the apparent horizon we use the Cactus thorn {\tt
ahfinder} \cite{ahfinder}.  We see that our results improve with
increased resolution for both the 2+1 and 3+1 simulations, as shown in
the upper subpanels.  For our
highest resolutions in the two cases, the deviation from the exact
value are smaller than 0.5\% and 2\%, respectively.  In the lower
subpanels, we show differences between results for varying numerical
resolutions, demonstrating second-order convergence until the horizon
finder begins to show errors at late times for our lowest-resolution runs.

\subsubsection{Moving Black Holes}\label{sec:movpunc}

We now turn to moving BH solutions.  In particular, we consider the
parameters discussed in detail in \cite{GodGauge}, a BH with linear
momentum $P_x=0.5{\mathcal M}$, traveling along the x-axis, starting
from an initial location $x_0=-3{\mathcal M}$.  For our initial data,
we calculate the extrinsic curvature and conformal factor from 
Eqs.~(\ref{eq:aijpunc}), (\ref{eq:confinit}), and (\ref{eq:upunc}) -- (\ref{eq:upunc2}).  We set the lapse initially to
$\alpha(t=0)=\psi^{-2}(t=0)$, and the shift to zero.  The spatial metric
is initially conformally flat.  According to
(\ref{eq:mass_translate}), the ADM mass of this configuration is
approximately $M_{\rm ADM} = 1.15625{\mathcal M}$, with the leading-order
error term appearing at order $P^4$.

We use unigrid simulations for easier comparison with the results of
\cite{GodGauge}, since the transformation to fisheye coordinates would
introduce new terms into the shift equations (namely
Eq.~(\ref{eq:shiftevolfish}) instead of Eq.~(\ref{eq:shiftevol2})).
Unigrid simulations are impractical in 3+1, so we perform axisymmetric
simulations instead.  We choose a numerical grid that extends twice as
far along the symmetry axis as it does radially.  We note that while
runs MP2b, MP2c, and MP2d, with numerical spacings ${\mathcal M}/24$,
${\mathcal M}/16$ and ${\mathcal M}/8$, respectively, use a grid that
extends to $r=\pm 70{\mathcal M}$ along the symmetry axis (along which
the BH moves) and $r=70{\mathcal M}$ along the radial axis, our highest
resolution calculation, run MP2a with spacing ${\mathcal M}/32$, only
extends to $r=55{\mathcal M}$ in both directions. 
We also perform 3+1 simulations of moving punctures, which do use
fisheye coordinates.  The parameters for these coordinates are the
same as in Sec.~\ref{sec:statbh}, except that we choose $R_1=6{\mathcal M}$ and
$R_2=9{\mathcal M}$ here to create a larger central region through which the
BH travels.  We evolve our moving puncture configurations
until $t_F =50{\mathcal M}$, during which time the BH remains inside the
central fisheye region.

\begin{figure}[ht!]
 \centering \leavevmode \epsfxsize=3in \epsfbox{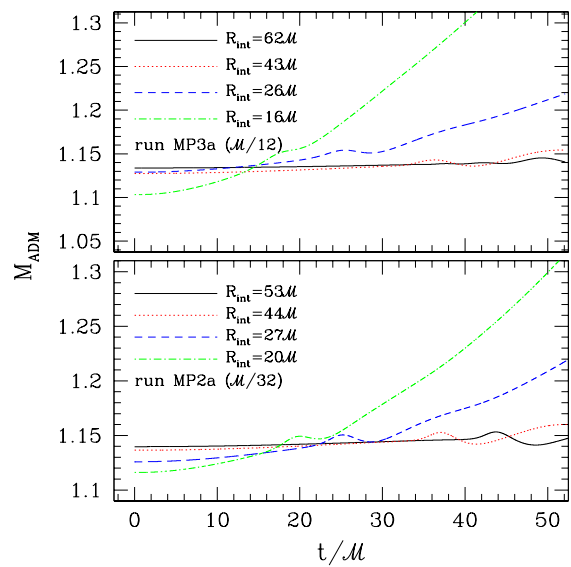}
 \caption{ADM mass versus time for moving puncture 3+1 evolution MP3a (see
Table \ref{table:vacruns}), measured using Eq.~(\protect\ref{eq:madm})
at surfaces of different radii.  In the top panel, we show results for
surfaces at constant fisheye radius $\bar{r}=8.0$ (dot-dashed), $9.2$
(dashed), $10.5$ (dotted), and $11.8$ (solid), which correspond to
physical radii $r_{\rm int}=16{\mathcal M}$, $26{\mathcal M}$,
$43{\mathcal M}$, and $62{\mathcal M}$, respectively.  In the bottom
panel, the integration surfaces are placed at $r_{\rm int}=20{\mathcal
M}$, $27{\mathcal M}$, $44{\mathcal M}$, and $53{\mathcal M}$ for 2+1
simulation MP2a.  For the outermost surfaces, we see deviations of less than
a percent over the course of the entire evolution in each case, just
as we did for the stationary puncture runs shown in
Fig.~\protect\ref{fig:statpunc_madm}.}
\label{fig:movpunc_madm}
\end{figure}

We show the ADM mass measured at various radii for
our highest resolution 3+1 and 2+1 simulations, MP3a and MP2a, in the
top and bottom panels of Fig.~\ref{fig:movpunc_madm}.  We again find
better agreement between the measured ADM mass and the exact value
when we evaluate Eq.~(\ref{eq:madm}) at larger radii. 

Conservation of the ADM mass for moving punctures is comparable to
that for stationary punctures, at approximately 1\% over a duration
$t=50{\mathcal M}$.  Some of the initial deviations for the ADM mass
in the 3+1 case, visible especially for the innermost surface, are a
result of small numerical errors associated with converting
$\bar{\tilde{\Gamma}}^i$ back into the physical value
$\tilde{\Gamma}^i$ through a coordinate transformation (see Appendix
\ref{app:fisheye}).

\begin{figure}[ht!]
 \centering \leavevmode \epsfxsize=3in \epsfbox{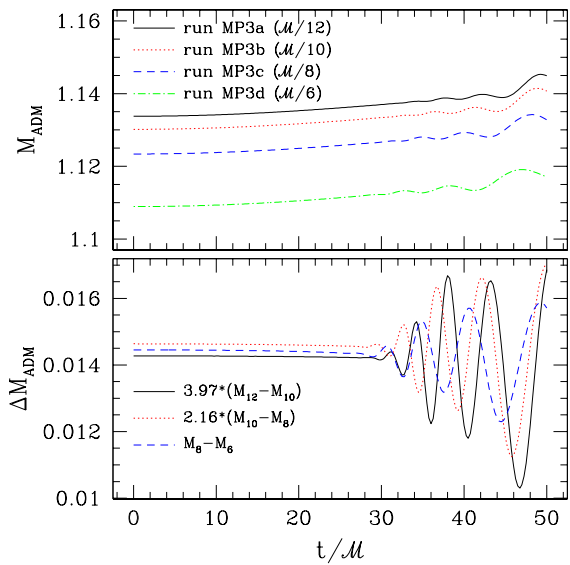}
 \caption{ADM mass versus time for moving puncture evolutions MP3 (see
 Table \ref{table:vacruns}), for runs with different numerical
 resolutions (top panel).  In each case, the integration surface was
 placed at $r=62{\mathcal M}$.  In the bottom panel, we show the
 differences in the measured ADM mass between pairs of ``neighboring''
 resolution, scaled to reflect the second-order convergence.}
\label{fig:movpunc_madmconv}
\end{figure}

As before, we confirm that the ADM mass converges to second order.  In
the top panel of Fig.~\ref{fig:movpunc_madmconv}, we show the ADM mass
measured at a surface of radius $r_{\rm int} = 62{\mathcal M}$ as a
function of time for our four 3+1 runs performed using different
numerical resolutions.  As
can be seen in Fig.~\ref{fig:movpunc_madmconv}, the values for the ADM
mass slowly drift to larger values.  We speculate that this is a
result of the constraints equations being solved only to order $P^2$,
which introduces a small but non-zero error.  Instead of testing
convergence to the approximate value of the ADM mass, we perform a
self-consistence convergence test by by computing the difference in
the ADM mass between pairs of runs in the bottom panel of the figure,
scaling the results appropriately in each case.  In terms of the
difference between lowest resolution pair, we scale the highest
resolution pair by a factor $(6^{-2}-8^{-2})/(10^{-2}-12^{-2})=3.97$
and the medium resolution pair by a factor
$(6^{-2}-8^{-2})/(8^{-2}-10^{-2})=2.16$.  Our findings suggest
second-order convergence throughout the evolution, even though at
times after t=30M the analysis is complicated by small-amplitude numerical
errors arising from discretization across fisheye transition regions.

Combining the results from Figs.~\ref{fig:movpunc_madm} and
\ref{fig:movpunc_madmconv}, we can extrapolate our results to both
asymptotic radii and infinite numerical precision.  With respect to
the radius, we take our results at $t=0$ measured at $r=43{\mathcal
M}$ and $62{\mathcal M}$ for run MP3a, and assume leading-order $1/r$
falloff behavior in the measured ADM mass.  With regard to numerical
resolution, we Richardson extrapolate using runs MP3a and MP3b
measured at $r=62{\mathcal M}$.  Combined, we find an extrapolated
value for the ADM mass of $M_{\rm ADM}=1.15615$, which falls within
$10^{-4}$ of the analytic value computed using Eq.~(\ref{eq:madm}) for
these approximate initial data.

\begin{figure}[ht!]
 \centering \leavevmode \epsfxsize=3in \epsfbox{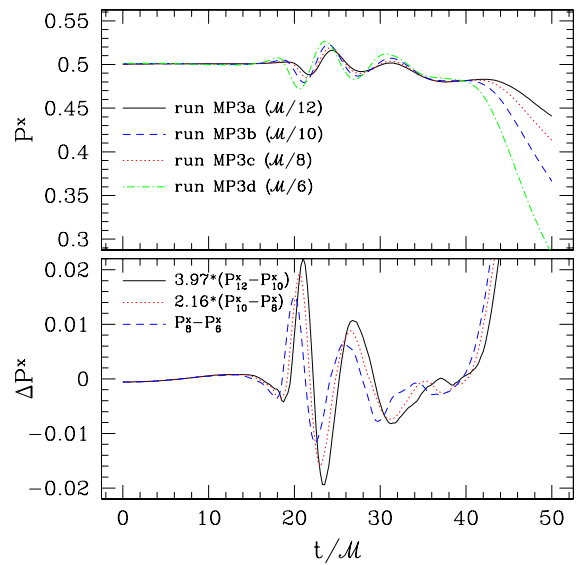}
 \caption{The top panel shows the linear momentum $P_x$
 (Eq.~(\protect\ref{eq:padm})) evaluated at $r_{int}=26{\mathcal M}$
 for our simulations MP3.  In the bottom panel, we show the difference
 between the values computed with different numerical resolutions,
 scaling the differences between the more accurate runs by factors of
 3.97 and 2.16, as in Fig.~\protect\ref{fig:movpunc_madmconv},
 to highlight the second-order convergence of our code.}
\label{fig:movpunc_p3d}
\end{figure}

For our 3+1 simulations we also compute the linear momentum from
Eq.~(\ref{eq:padm}).  In the top panel of Fig.~\ref{fig:movpunc_p3d},
we show the linear momentum as a function of time, calculated at
surfaces placed at radii $r_{\rm int}=26.8{\mathcal M}$.
In general, we see excellent agreement with the exact analytic value
up until $t\sim r_{\rm int}$, for the reasons noted above.  
Recall that we adopt initial data for
moving BHs that are analytic but approximate, since they
solve the constraint equations only to order $P^2$.  The resulting
error represents ``junk'' that propagates from the strong-field region
outwards, and reaching the integration surface $r_{\rm int}$ at
approximately $t\sim r_{\rm int}$.  Eq.~(\ref{eq:padm}) for the
linear momentum also assumes that the coordinate vector pointing in the
direction of the momentum is a Killing vector of the asymptotic,
conformally related metric (see Appendix A of \cite{Cook94}); this
assumption breaks down once the ``junk'' reaches $r_{\rm int}$.
Accordingly, evaluating Eq.~(\ref{eq:padm}) leads to an error in the
linear momentum, as is evident from Fig.~\ref{fig:movpunc_p3d}.
In the bottom panel of
the figure, we show the convergence behavior of the linear momentum.
As for the ADM mass we find larger error terms once numerical errors
reach the integration surface $r_{\rm int}$.

\begin{figure}[ht!]
 \centering \leavevmode \epsfxsize=3in \epsfbox{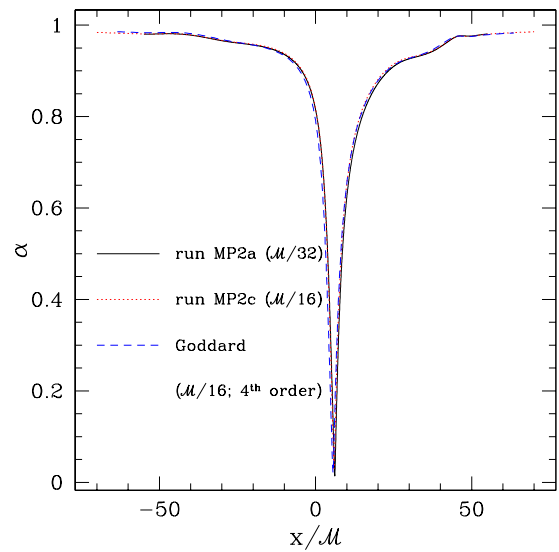}
 \caption{Lapse function $\alpha$ shown at $T=40{\mathcal M}$ along
 the path traveled by the moving puncture, for our axisymmetric 2+1
 runs with numerical resolutions of ${\mathcal M}/32$ (run MP2a;
 dashed) and ${\mathcal M}/16$ (run MP2c; dotted), along with the
 numerical results of GSFC (solid) as discussed in
 \protect\cite{GodGauge}.  The code used for the latter has numerical
 resolution ${\mathcal M}/16$, but uses fourth-order differencing,
 whereas we use a second order scheme.}
\label{fig:godcompare}
\end{figure}

As a further test we compare our numerical results for the lapse
function $\alpha$ in runs MP2a and MP2c with those obtained by the
numerical relativity group at NASA's Goddard Space Flight Center
(\cite{GodGauge}, hereafter GSFC).  In Fig.~\ref{fig:godcompare} we
graph $\alpha$ along the trajectory of the puncture at $t =
40{\mathcal M}$.  The simulations of GSFC was performed using a
numerical resolution of ${\mathcal M}/16$, but features a fourth-order
accurate differencing scheme.  We nevertheless find excellent
agreement between our results.

\begin{figure}[ht!]
 \centering \leavevmode \epsfxsize=3in \epsfbox{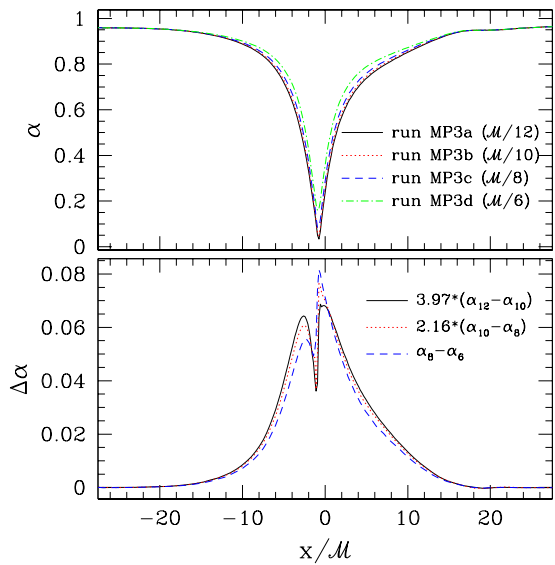}
 \caption{The top panel shows the lapse function $\alpha$ measured on
 the x-axis, along which the puncture travels, at $t=20{\cal M}$. 
 In the bottom panel, we show the difference
 between the values computed with different numerical resolutions,
 rescaling the differences as in Fig.~\protect\ref{fig:movpunc_madmconv},
 to highlight the second-order convergence of our code.}
\label{fig:movpunc_lapconv}
\end{figure}

Extending our tests to the 3+1 moving puncture cases, we show in the
top panel of Fig.~\ref{fig:movpunc_lapconv} the lapse function along
the axis on which the BH travels, at $t=20{\mathcal M}$.  In the bottom panel, we
show the convergence of the lapse function by taking the difference
between the runs and rescaling to second-order, finding good agreement
throughout. The slight discrepancy near the puncture position
is due to the fact that there is a sharp trough in the lapse function
there, but the position of the puncture itself falls at slightly 
different coordinate positions for the three runs. Indeed,
since the shift vector, like all field quantities, 
is second-order convergent in the
numerical resolution, so too is the speed at which the puncture moves, since
$dx^i/dt=-\beta^i(x^i)$.  

\begin{figure}[ht!]
 \centering \leavevmode \epsfxsize=3in \epsfbox{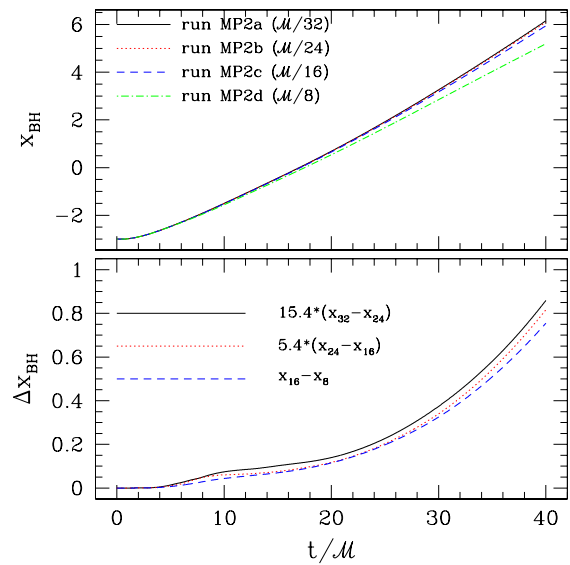}
 \caption{Position of the BH puncture versus time for our 2+1
 axisymmetric runs with numerical resolutions of ${\mathcal M}/32$
 (run MP2a; solid), ${\mathcal M}/24$ (run MP2b; dotted), ${\mathcal
 M}/16$ (run MP2c; dashed), and ${\mathcal M}/8$ (run MP2d;
 dot-dashed).  In the bottom panel, we show the difference in position
 versus time for the same runs, noting that we see second order
 convergence in the position of the puncture.}
\label{fig:movpunc_x}
\end{figure}

Turning our attention to the position of the puncture, we show in
Fig.~\ref{fig:movpunc_x} the position of the BH puncture
versus time in our three axisymmetric calculations performed at
different numerical resolutions, along with the Richardson
extrapolation value.  We see good agreement, and note that numerical
errors associated with coarser resolutions slow down the BH away from
the proper asymptotic velocity.  In the bottom panel of the figure, we
show the difference in position versus time for our 2+1 runs, again
suggesting second-order convergence.

\section{Matter Tests}
\label{sec:matter}

In this Section we describe relativistic hydrodynamics simulations in
the presence of puncture BHs.  We are particularly interested
in testing how the accretion of matter onto BHs can be
simulated within the moving puncture approach.  The exact
Bondi solution for accretion onto a static Schwarzschild BH provides a perfect test bed for
these purposes.  A detailed discussion of the relativistic Bondi
solution can be found in Appendix~G of \cite{ShapTeuk}; we summarize
all relevant expressions in Appendix \ref{app:bondi}.  In Section
\ref{sec:statbondi} we test our code's capability of simulating the
accretion onto a static BH; in Section \ref{sec:movbondi} we
treat the identical problem, but viewed in a frame in which the BH is
moving and represented by a moving puncture BH.  We summarize our Bondi simulations in Table
\ref{table:matruns}.

\begin{table}
\caption{Summary of our matter evolution simulations.  Key equations
for the static Bondi solution, denoted by ``SB'', are summarized in
Appendix \ref{app:bondi}.  Moving Bondi solutions, constructed
as described in Appendix \ref{app:bondi_boost}, are
``boosted'' to have a linear velocity $v^x = 0.1$ 
}
\begin{tabular}{l|l|l|l|l}
\hline
Name & Initial Data & Grid & Resolution & $t_F$ \\
\hline\hline
\multicolumn{5}{c}{2+1 w/Fisheye; $|\bar{x}_{fish}^i|\le 15{\mathcal M}$, $|x^i|\le 136{\mathcal M}$}\\
\hline
SB2a & Static Bondi & $360\times360$ & ${\mathcal M}$/24 & 100${\mathcal M}$\\
SB2b & Static Bondi & $240\times240$ & ${\mathcal M}$/16 & 100${\mathcal M}$\\
SB2c & Static Bondi & $120\times120$ & ${\mathcal M}$/8 & 100${\mathcal M}$\\
\hline
\multicolumn{5}{c}{3+1 w/Fisheye; $|\bar{x}_{fish}^i|\le 10{\mathcal M}$, $|x^i|\le 56{\mathcal M}$}\\
\hline
SB3a & Static Bondi & $240^2\times 120$ & ${\mathcal M}$/12 & 100${\mathcal M}$\\
SB3b & Static Bondi & $200^2\times 100$ & ${\mathcal M}$/10 & 100${\mathcal M}$\\
SB3c & Static Bondi & $160^2\times 80$ & ${\mathcal M}$/8 & 100${\mathcal M}$\\
SB3d & Static Bondi & $120^2\times 60$ & ${\mathcal M}$/6 & 100${\mathcal M}$\\
\hline
\multicolumn{5}{c}{3+1 w/Fisheye; $P^i=0.1$, $|\bar{x}_{fish}^i|\le 10{\mathcal M}$, $|x^i|\le 56{\mathcal M}$}\\
\hline
MB3a & Moving Bondi & $240^2\times 120$ & ${\mathcal M}$/12 & 100${\mathcal M}$\\
MB3b & Moving Bondi & $200^2\times 100$ & ${\mathcal M}$/10 & 100${\mathcal M}$\\
MB3c & Moving Bondi & $160^2\times 80$ & ${\mathcal M}$/8 & 100${\mathcal M}$\\
MB3d & Moving Bondi & $120^2\times 60$ & ${\mathcal M}$/6 & 100${\mathcal M}$\\
\hline
\end{tabular}
\label{table:matruns}
\end{table}

\subsection{Stationary Bondi solutions}
\label{sec:statbondi}

The analytic solution describing Bondi accretion onto a stationary
BH is usually given in Schwarzschild coordinates (see, e.g.,
Appendix~G of \cite{ShapTeuk} for a detailed description, and Appendix
\ref{app:bondi_equations} for a summary).  To construct initial data
for our dynamical simulations, we transform this solution into
isotropic coordinates, as described in Appendix \ref{app:bondi_iso}.
Since isotropic coordinates become singular on the horizon at
$r={\mathcal M/2}$, so does
the fluid velocity when expressed in these coordinates.  We therefore
adjust the fluid initial data artificially in the immediate vicinity
of the BH.  Specifically, for the rest-mass density, we fit a
quadratic function between $r={\mathcal
M}/2$ and $r={\mathcal M}$ so that its radial derivative matches the
analytic Bondi solution at $r={\mathcal M}$ and the derivative goes to 
zero at $r={\mathcal M}/2$.  We note that for stationary BHs,
${\mathcal M}=M_{\rm ADM}$.  Inside the
horizon at $r={\mathcal M}/2$, we set the density equal to a small
positive value at the origin, plus a term with radial
dependence $\propto 1-\cos(2\pi r/{\mathcal M})$ used to establish a
smooth fit.  For the velocity, we simply set $u =\left.u \right|_{r =
{\mathcal M}} \times (r/{\mathcal M})$ for $r <{\mathcal M}$.  Since the flow
is supersonic and directed inward, this does not affect the exterior
solution, and even within $r<{\mathcal M}$ the fluid solution quickly
settles into an equilibrium flow as our spacetime slicing evolves
towards the late-time solution that penetrates the horizon smoothly.

To match to the stationary Bondi solution, in which the self-gravity
of the gas is negligible, we require that the mass accretion
rate multiplied by the integration time -- which we choose to be $t_F
= 100 {\mathcal M}$ -- remain small with respect to the mass of the
BH.  Thus, we set $\dot{M}=10^{-4}$ for all runs shown here.  
We set the sonic areal radius to $r_s=10{\mathcal M}$.  
The proper infall time required for the fluid to travel from
$r\approx 9{\mathcal M}$ ($r_s$ in isotropic radii) to the horizon at
$r_h={\mathcal M}/2$ is $23{\mathcal M}$, so that we evolve for just
over four freefall times.  The gas is at rest at infinity, with a
uniform density of $\rho_\infty=6.2\times 10^{-8}$.  The polytropic
index is chosen to be $n=3$ (thus, the adiabatic index is $\Gamma=4/3$).

For the gauge conditions used in moving puncture simulations, the
Bondi solution is not time-independent.  Similarly to the vacuum
solutions described in Section \ref{sec:vacuum}, the evolution passes
through a transient, time-dependent phase, and then settles down into
a new equilibrium.  This new equilibrium solution describes 
the usual Bondi solution, but expressed in a different
coordinate system.  To analyze this solution and compare it with the
analytical solution (which is given in Schwarzschild coordinates), we
therefore need to compare invariants.

One such invariant is the rate of change of the fluid rest density $\rho_0$
as measured by an observer moving with the fluid,
\begin{equation} \label{eq:dotrho}
\dot{\rho_0}\equiv d\rho_0/d\tau\equiv u^\mu\partial_\mu \rho_0,
\end{equation}
where $u^\mu$ is the fluid's 4-velocity. 

\begin{figure}[ht!]
 \centering \leavevmode \epsfxsize=3in \epsfbox{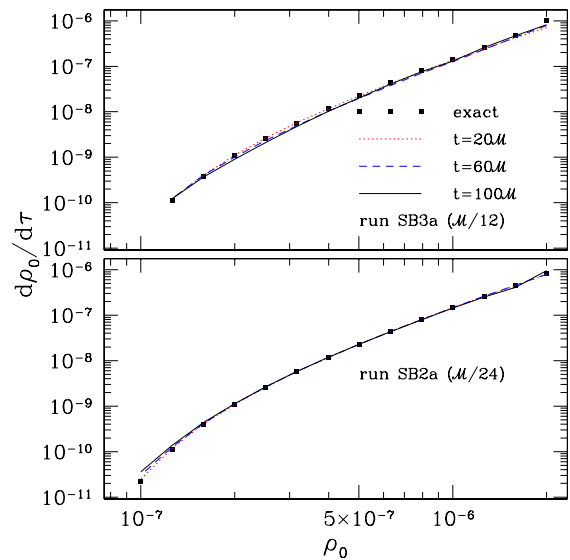}
 \caption{Time rate of change of the rest-mass
 density, $d\rho_0/d\tau$, as a function of the rest-mass density.  
We show results for our highest resolution
 3+1 (top panel) and 2+1 (bottom panel) calculations.  The exact
 solution is represented by points, along with our profiles at
 $t=20{\mathcal M}$ (dotted curve), $t=60{\mathcal M}$ (dashed), and
 $t=100{\mathcal M}$ (solid).}
\label{fig:statbondi_rhodot}
\end{figure}

In Fig.~\ref{fig:statbondi_rhodot}, we show $d\rho_0/d\tau$ measured
along the z-axis (perpendicular to the symmetry axis) 
as a function of the rest-mass density itself.  
In the top panel, we show
the results for our highest resolution 3+1 simulation, run SB3a with
${\mathcal M}/12$ spacing, whereas in the bottom panel we plot values
for run SB2a (${\mathcal M}/24$).  In both cases we plot the exact
solution as points, along with our numerical profiles at
$t=20{\mathcal M}$, $t=60{\mathcal M}$, and $t=100{\mathcal M}$.  We
find good agreement throughout the evolution with variations of no
more than $10\%$ and $4\%$ respectively for the 3+1 and 2+1
simulations.

\begin{figure}[ht!]
 \centering \leavevmode \epsfxsize=3in \epsfbox{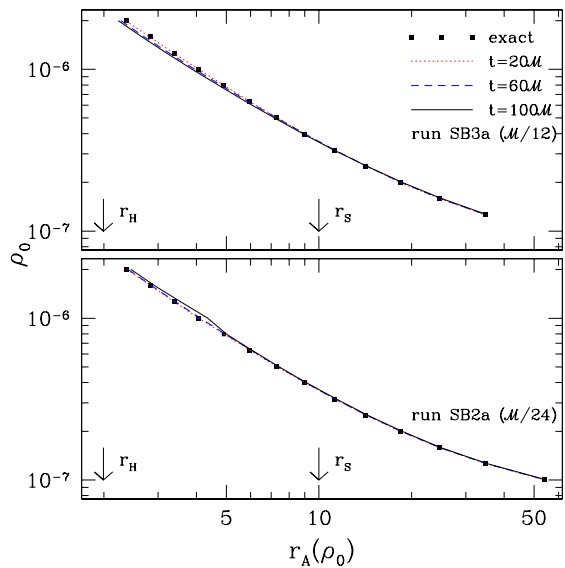}
 \caption{ Rest-mass density as a function of the areal radius of the
 corresponding isodensity surface, shown on a log-log scale.  We show
 both our highest resolution 3+1 run (run SB3a; top panel) and 2+1 run
 (run SB2a; bottom panel).  The density values at which we compute the
 surfaces are the same as in Fig.~\protect\ref{fig:statbondi_rhodot},
 as are all other conventions, though we note that the density axis
 have been flipped.  We mark the BH horizon $r_h$ and the sonic point
 of the flow $r_s$ with arrows, noting that we can see evidence for
 the well-known phase transition in the density-radius relation at the
 sonic point.}
\label{fig:statbondi_rhoarea}
\end{figure}

As an additional test we compute the average areal radius of
isodensity surfaces, defined by
\begin{equation} \label{eq:r_A}
r_A\equiv \left(\frac{{\mathcal A}}{4\pi}\right)^{1/2},
\end{equation}
in terms of the surface's proper area ${\mathcal A}$.  Evidently, this
is again a coordinate-independent quantity.  For the essentially spherically
symmetric isodensity surfaces considered in this Section the average radius
must of course be equal to its local value, but the definition
Eq.~(\ref{eq:r_A}) generalizes to the non-spherical configurations in Section
\ref{sec:movbondi}.  In Fig.~\ref{fig:statbondi_rhoarea}, we show the
average radii of the same isodensity surfaces described in
Fig.~\ref{fig:statbondi_rhodot}, following the same conventions.  We
note that it is possible to spot the phase transition between subsonic
and supersonic behavior at the sonic radius $r_S$, marked by an arrow,
which causes a shift in the power-law index of the radius-density
relation.  We also marked the horizon at $r_h$ with a second arrow.
Again, we see that the results are stable for a long period, with
variations of no more than $5\%$ and $3\%$, respectively.

\begin{figure}[ht!]
 \centering \leavevmode \epsfxsize=3in \epsfbox{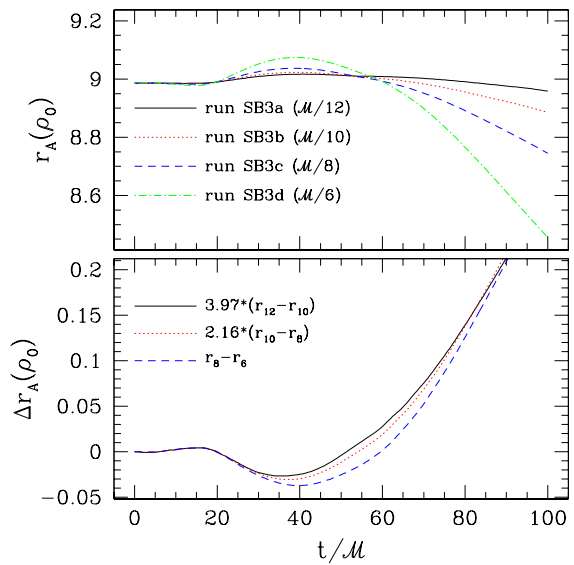}
 \caption{The average areal radius of an isodensity surface with
   density $\rho_0=3.8\times 10^{-7}$ as a function of time for 3+1
   runs with different numerical resolutions: runs SB3a (${\mathcal
   M}/12$; solid), SB3b (${\mathcal M}/10$; dotted), SB3c (${\mathcal
   M}/8$; dashed), and SB3d (${\mathcal M}/6$; dot-dashed). In the
   bottom panel, we show the pairwise differences between runs.}
\label{fig:statbondi_areaconv}
\end{figure}

To test the convergence of our implementation of relativistic
hydrodynamics we have performed Bondi evolution calculations using the
same set of numerical resolutions used previously for the vacuum
puncture calculations described in Sec.~\ref{sec:advect}.  In
Fig.~\ref{fig:statbondi_areaconv}, we show $r_A$ for
runs with differing numerical resolutions at the same fixed density
value, chosen to be $\rho_0=3.8\times 10^{-7}$, which lies near the
center of our logarithmic range and corresponds to a location close to
the sonic radius.
We see that matter variables converge to second order in the grid resolution,
just as the field variables do.  In Fig.~\ref{fig:statbondi_areaconv},
we note that the extrapolated solution does seem to expand slowly over
time, but that this effect represents approximately a $1\%$ change
over a period of $t=100{\mathcal M}$.  This effect is
caused by the presence of the outer boundary.  At sufficiently early
times, our numerical solution converges to the analytical solution in regions
that are causally disconnected from the outer boundary.  

\begin{figure}[ht!]
 \centering \leavevmode \epsfxsize=3in \epsfbox{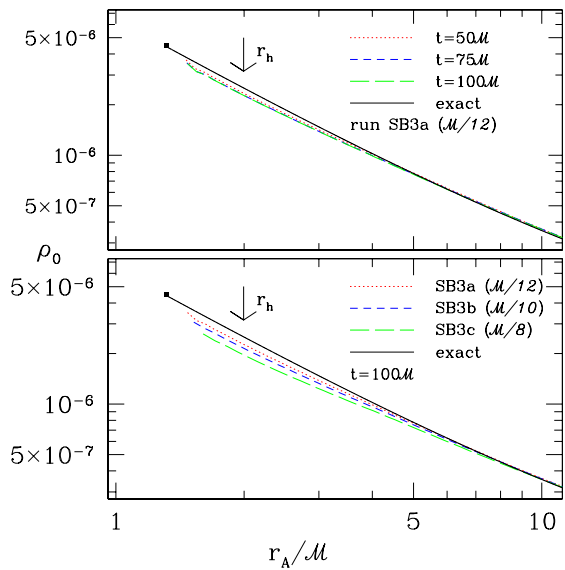}
 \caption{The top panel shows the rest-mass density
     $\rho_0$ as a function of areal radius $r_A$ for run SB3a at
     times $t=50{\mathcal M}$ (dotted), $t=75{\mathcal M}$ (dashed),
     and $t=100{\mathcal M}$ (dot-dashed), as well as the exact Bondi
     solution (solid). We exclude the three innermost gridpoints from
     each run, which are directly affected by our hydrodynamical
     ``fixes''.  We see the solution remains smooth and accurate
     across the horizon, at $r_A=r_h=2{\mathcal M}$, which is marked with
     an arrow, remaining finite everywhere and nearly constant in
     time. In the bottom panel of the figure we show results at
     $t=100{\mathcal M}$ for runs with three different numerical
     resolutions, showing the expected convergence toward the exact
     solution.}
\label{fig:statbondi_rhor}
\end{figure}

As we discussed in Section \ref{sec:movpuncevolve}, the point $r = 0$
corresponds to a surface of finite, positive areal radius (see
\cite{JenaGeo1,JenaGeo2}), and represents a coordinate singularity
only.  Matter reaching this point therefore represents matter crossing
a surface of a certain finite areal radius inside the horizon.
Accordingly, all fluid quantities should remain finite at $r=0$.  In the top
panel of Fig.~\ref{fig:statbondi_rhor}, we show the rest-mass density
as a function of areal radius for our 
highest-resolution 3+1 stationary Bondi result at different times,
along with the analytical solution.  We eliminate from the figure the
three innermost grid points, where the hydrodynamical ``fix'' we apply
(recall Sec.~\ref{sec:hydro})
directly affects the values of the primitive hydrodynamical variables
computed from the conserved set.  We see that the flow extends
smoothly within the horizon, extending inward to nearly the asymptotic
limiting value of $r_A=1.31{\mathcal M}$ \cite{JenaGeo1}.  Thus, the
matter maintains a regular flow pattern into the BH, as we would
expect, remaining finite and well-behaved indefinitely.  In the bottom
panel of the figure, we show results at $t=100{\mathcal M}$ for runs
of varying resolution, showing that we converge toward the analytic
solution as we increase our numerical resolution, and that the
physical region affected by the hydrodynamical fixes decreases in size
as the resolution increases.  We find the convergence is second-order
at larger radii, and approximately first-order nearer the puncture
where differencing errors across the puncture and the
hydrodynamical fixes impose small-scale oscillations in the density.

\subsection{Bondi solutions for moving punctures}
\label{sec:movbondi}

We would also like to study our code's ability to handle matter in the
presence of a moving puncture.  For these purposes we
study the Bondi solution as viewed in a frame in which the
Schwarzschild BH puncture is moving.
Our method is described in Appendix \ref{app:bondi_boost}.

\begin{figure}[ht!]
 \centering \leavevmode \epsfxsize=3in \epsfbox{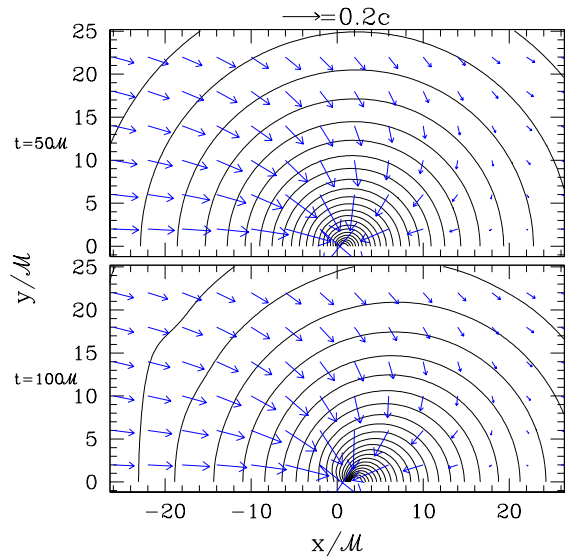}
 \caption{Contours of the rest-mass density for our moving Bondi
 evolution run MB3a, shown as slices through the equatorial plane at
 $t=50$ (top panel) and $t=100{\mathcal M}$ (bottom panel). Our
 simulation extends to cover the negative $y$-plane as well, but is
 visually indistinguishable from being completely symmetric.  Density
 contours begin at $\rho_0=10^{-7}$, and are spaced logarithmically,
 16 per decade.  A velocity vector representing a magnitude $v=0.2c$
 is shown above the figure for reference.}
\label{fig:movbondi_contour}
\end{figure}

In this Section we consider BHs with a velocity of $v^x\equiv
P^x/{\mathcal M} =
0.1$, and let the BH start at a coordinate location of
$x=-2.5{\mathcal M}$.  As in the stationary Bondi case discussed in
Sec.~\ref{sec:statbondi} above, we evolve our 3+1
calculations for a duration $t=100{\mathcal M}$, equivalent to
approximately 4 sonic radius freefall times, at which point the BH has
moved to a coordinate location of $x=1.8{\mathcal M}$.  To visualize
the resulting evolution, we show density contours with overlaid arrows
representing the velocity field at $t=50{\mathcal M}$ and
$100{\mathcal M}$ for run MB3a in Fig.~\ref{fig:movbondi_contour}.  We
see a clear pattern of translation as the entire solution evolves.

\begin{figure}[ht!]
 \centering \leavevmode \epsfxsize=3in \epsfbox{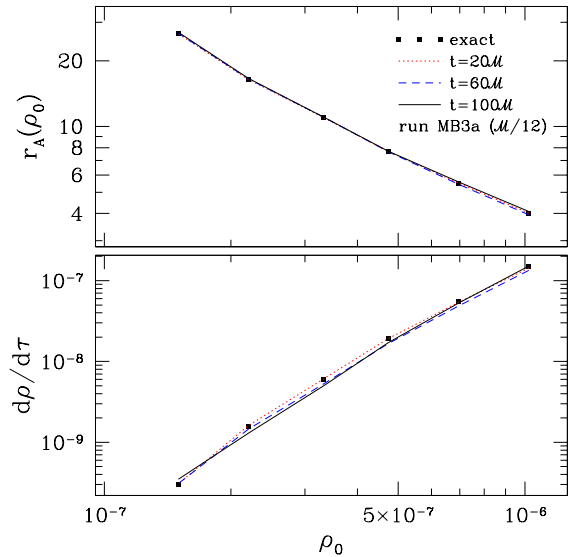}
 \caption{Average proper radius of isodensity surfaces $r_A$ (top
 panel) and time rate of change of the density along the
 z-axis, $d\rho_0/d\tau$, (bottom panel) shown as a function of the
 rest mass density on a log-log scale for run MB3a.  The exact
 solutions are shown as square points.  Results are shown at
 $t=20{\mathcal M}$ (dotted), $t=60{\mathcal M}$ (dashed), and
 $t=100{\mathcal M}$ (solid).}
\label{fig:movbondi_arearhodot}
\end{figure}

\begin{figure}[ht!]
 \centering \leavevmode \epsfxsize=3in \epsfbox{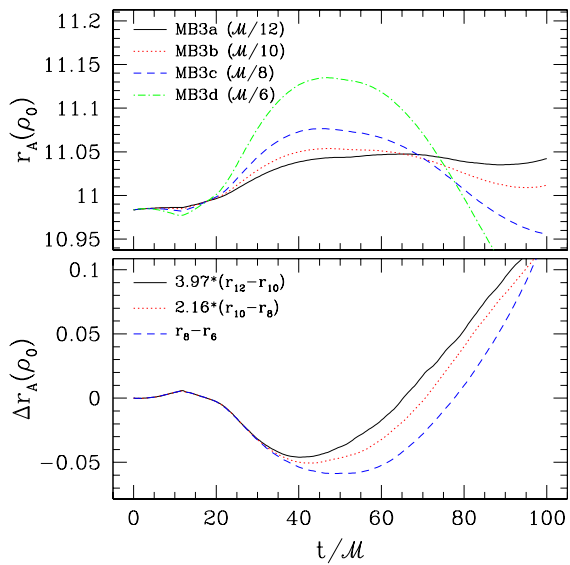}
 \caption{The average areal radius $r_A$ of the isodensity surface
 with density $\rho_0=3.8\times 10^{-7}$ as a function of time (top
 panel) for runs with different numerical resolutions: runs MB3a
 (${\mathcal M}/12$; solid), MB3b (${\mathcal M}/10$; dotted), MB3c
 (${\mathcal M}/8$; dashed), and MB3d (${\mathcal M}/6$; dot-dashed).
 In the bottom panel, we show the scaled pairwise differences between
 the runs.}
\label{fig:movbondi_areaconv}
\end{figure}

In Fig.~\ref{fig:movbondi_arearhodot} we show $r_A$ as in
Eq.~(\ref{eq:r_A}), and $d\rho_0/d\tau$ from Eq.~(\ref{eq:dotrho}) as
a function of density $\rho_0$ for simulation MB3a, our highest
resolution moving Bondi run (compare Figs.~\ref{fig:statbondi_rhodot}
and \ref{fig:statbondi_rhoarea} for stationary Bondi solutions).  As
before, the exact solutions are given by the points.  We find that
$r_A$ agrees with the analytical solution to within about $3\%$.  The
co-moving time derivative of the density $d \rho_0/d\tau$, on the
other hand, shows larger deviations of up to $20\%$.  In
Fig.~\ref{fig:movbondi_areaconv}, we show the convergence behavior in
the average areal radius of the isodensity surfaces as a function of
time for our runs, following the same conventions as
Fig.~\ref{fig:statbondi_areaconv}.  We see the same convergent
behavior as in the vacuum and stationary Bondi cases: second-order
convergence followed by the appearance of higher-order correction
terms at approximately $t=40{\mathcal M}$.

\section{Discussion and Future Calculations}\label{sec:discussion}

In this paper we have performed several test simulations involving the
modeling of BHs within the moving puncture approach, both in vacuum
and in the presence of a relativistic fluid.

Our vacuum tests focus on evolutions of both stationary and moving
BHs.  We find that the code is
second-order convergent, as expected, but is limited primarily by the
maximum numerical resolution we can achieve with our current second-order
formulation (cf. the ADM mass convergence
demonstrated in \cite{Marro1} at finer numerical resolution). 
To remedy this issue, we have introduced fourth-order
spatial differencing into our code, results of which will be reported
in \cite{EFLSB}.
We demonstrate that we can reproduce
numerically the analytical solution of \cite{BaumPunc} for an isolated
stationary BH, in line with previous studies of stationary isolated
punctures (see, e.g., \cite{Jena3,Brown,Brown2}).
We also demonstrate that we can artificially modify the initial data
in the BH interior, and even violate the constraints there, and still settle
down to the same late-time asymptotic solution without significantly 
affecting the
evolution in the BH exterior.  While this result is not surprising, it
is reassuring, and also has some implications for future simulations,
as we discuss below.

To test the ability of the moving puncture approach to handle the
flow of matter onto BHs, we study relativistic Bondi flow
both for stationary and moving BHs and again compare with the
analytic solution.  Our findings demonstrate that all fluid
variables remain regular throughout and do not need excision.  This
result can be understood in terms of the studies of
\cite{JenaGeo1,JenaGeo2}, who demonstrate that the puncture
represents a limiting surface of finite areal radius, and hence a
coordinate singularity only.  Moving puncture evolutions never
encounter the BH's central spacetime singularity, and cover regular
regions of the spacetime only.  Hence all fluid invariants are regular
throughout the puncture BH interior.

This paper represents a stepping-stone in our group's efforts to model
relativistic BHNS binaries.  We have previously constructed BHNS initial
data, using the conformal thin-sandwich formalism and imposing
quasi-equilibrium boundary conditions on the BH's horizon (see
\cite{TBFS1,TBFS2}).  We have also performed preliminary dynamical
simulations in conformal gravitation (see \cite{FBSTR,FBST}).  We now
plan to adopt the puncture
method to simulate BHNS binaries fully self-consistently, similar
to the calculations of \cite{SUpunc1,SUpunc2}.

In contrast to \cite{SUpunc1,SUpunc2}, we plan to evolve
quasi-equilibrium conformal thin-sandwich 
initial data.  This poses the conceptional problem
that the initial data use excision to model the BH, while the
dynamical evolution requires initial
data everywhere.  As shown above, however, our experiments with
single BHs demonstrate that we can replace the initial data in
the BH interior with arbitrary sufficiently smooth functions without
significantly affecting the
late-time solution, or the evolution, in the BH exterior.  In
fact, this suggests that moving puncture simulations of BHBH binaries
may also use initial data constructed in the conformal thin-sandwich
formalism (see, e.g., \cite{CPInit,Caudill}),
which are believed to represent true quasi-equilibrium configurations
more accurately \cite{BIW}.  The evolution of conformal thin-sandwich
binary BH initial data will be considered in detail in a forthcoming
paper \cite{EFLSB}.

\acknowledgments

It is a pleasure to thank Greg Cook, James van Meter, Carlos Lousto,
Yuk Tung Liu, and Hwei-Jang Yo for helpful conversations.  JAF is
supported by an NSF Astronomy and Astrophysics Postdoctoral Fellowship
under award AST-0401533.  This work was supported in part by NSF
grants PHY-0205155 and PHY-0345151 and NASA Grants NNG04GK54G and
NNX07AG96G to the
University of Illinois, and NSF Grant PHY-0456917 to Bowdoin College.
\begin{appendix}

\section{Fisheye Coordinates}\label{app:fisheye}

Fisheye coordinates are defined through a purely radial coordinate
transformation, in which we define a fisheye radius $\bar{r}$ in terms
of the physical radius $r$ according to
\begin{equation}
\bar{r}=a_nr+\sum_{i=1}^{n}\frac{(a_{i-1}-a_i)s_i}{2\tanh(R_i/s_i)}
\ln\left(\frac{\cosh((r+R_i)/s_i)}{\cosh((r-R_i)/s_i)}\right),
\end{equation}
(see Eqs.~(3) and (4) of \cite{UTB2}).  Here the $a_i$ coefficients
determine the ``stretching'' of the radial coordinate in several
different regions, labeled by the $i$'s, the $R_i$'s define the center
of transition zones between these regions, and the $s_i$'s determine
the width of these transition zones.  The derivative of this
expression is given by
\begin{eqnarray}
\frac{d\bar{r}}{dr}&=&a_n+\sum_{i=1}^n\frac{a_{i-1}-a_i}{\tanh(R_i/s_i)}\times
\\
&&\left(\frac{\tanh((r+R_i)/s_i)-\tanh((r-R_i)/s_i)}{2}\right) \nonumber,
\end{eqnarray}
which helps to understand how this transformation works.  Assume that
we arrange the $R_i$ terms in such a way that neighboring transition
zones do not overlap, i.e.~$R_i-s_i > R_{i-1}+s_{i-1}$.  Outside of the
transition zones, i.e.~for radii $R_{m-1}+s_{m-1}<r<R_m-s_m$, the
derivative $d\bar{r}/dr$ is then given approximately by that region's
coefficient $a_{m-1}$.  This is because the last term in parentheses takes a value approximately equal
to zero for all $i\le m-1$ and approximately one for $i\ge m$.  The fisheye
transitions act in many ways like an effective fixed-mesh refinement,
with spherical transitions separating regions with different
resolutions.  Angles with respect to the origin remain unchanged by
the transformation, and spheres centered on the origin transform into
spheres, albeit with a different radius.  We always apply the fisheye
transformation in terms of the origin of our coordinates, regardless
of the position of the BH in cases where it is moving across the grid.

The fisheye transformation is purely radial.  For Cartesian
coordinates we define
\begin{equation}
\bar{x}^i=x^i \left(\frac{\bar{r}}{r}\right).  
\end{equation}
The Jacobian of the coordinate transformation and its inverse are
given by
\begin{eqnarray}
  \frac{\partial x^i}{\partial \bar{x}^j} &=& \delta^i{}_j
\left( \frac{r}{\bar{r}}\right) + \frac{\bar{x}^i \bar{x}^j}{\bar{r}^2}
\left( \frac{dr}{d\bar{r}}-\frac{r}{\bar{r}}\right) \ , \\
  \frac{\partial \bar{x}^i}{\partial x^j} &=& \delta^i{}_j
\left( \frac{\bar{r}}{r}\right) + \frac{\bar{x}^i \bar{x}^j}{\bar{r}^2}
\left( \frac{d\bar{r}}{dr}-\frac{\bar{r}}{r}\right) \ .
\end{eqnarray}

To construct initial data on a fisheye grid, the most direct method is
to evaluate all quantities at the physical coordinates represented by
the point, followed by an appropriate coordinate conversion.  Since
the transformation is purely spatial, all time-components of
4-vectors remain unaffected, and we can restrict the transformation
to spatial components only, e.g.
\begin{equation}
\bar{v}^i= \frac{\partial \bar{x}^i}{\partial x^j}v^j;~~\bar{v}_i=
\frac{\partial x^j}{\partial \bar{x}^i}v_j,
\end{equation}
and
\begin{equation}
 \bar{\gamma}_{ij} = \frac{\partial x^l}{\partial \bar{x}^i}
\frac{\partial x^m}{\partial \bar{x}^j} \gamma_{lm} \ .
\end{equation}
The rest-density, for example, is the time-component of the fluid's
density 4-vector, and is hence invariant under these
transformations.

The determinant of the Jacobian is given by
\begin{equation}
J = \det(\partial x^i/\partial \bar{x}^j) = 
\left( \frac{r}{\bar{r}}\right)^2 \frac{dr}{d\bar{r}}.
\end{equation}
We then have
\begin{equation}
  \bar{\gamma} \equiv \det(\bar \gamma_{j}) = J^2 \gamma,
\end{equation}
indicating that the determinant of the metric is a tensor density of
weight 2.  Using a conformal transformation $\gamma_{ij} = \psi^4
\tilde \gamma_{ij}$ so that the determinant of the conformally related
metric is unity, $\tilde \gamma = 1$, implies
\begin{equation}
\bar{\psi}^{12}= J^2 \psi^{12},
\end{equation}
meaning that the conformal factor is a tensor density of weight $1/6$.
The conformal factor and its logarithm $\phi = \ln \psi$ then
transform according to
\begin{eqnarray}
 \bar{\psi}&=&\left(\frac{r}{\bar{r}}\right)^{1/3}\left(
\frac{dr}{d\bar{r}}\right)^{1/6} \psi, \\ \bar{\phi}\equiv
\ln\bar{\psi}&=&\frac{1}{3}\ln\left(\frac{r}{\bar{r}}\right)+\frac{1}{6}\ln\left(
\frac{dr}{d\bar{r}}\right)+\ln\psi.
\label{eq:phibar}
\end{eqnarray}
Our conformal field quantities, $\tilde{\gamma}_{ij}$ and
$\tilde{A}_{ij}$, both transform according to the relation,
\begin{equation}
\bar{\tilde{\gamma}}_{ij} = \left(\frac{r}{\bar{r}}\right)^{4/3}\left( \frac{dr}{d\bar{r}}\right)^{2/3}\frac{\partial x^l}{\partial \bar{x}^i}
\frac{\partial x^m}{\partial \bar{x}^j} \tilde{\gamma}_{lm} \ .
\end{equation}
The most complicated transformation is that of $\tilde{\Gamma}^i$,
given by 
\begin{equation}
\bar{\tilde{\Gamma}}^j
= J^{2/3} \frac{\partial \bar{x}^j}{\partial x^l}\tilde{\Gamma}^l-
\tilde{\gamma}^{ln}\left[
-\frac{1}{2}\frac{\partial J^{2/3}}{\partial x^n}
\frac{\partial  \bar{x}^j}{\partial x^l}+
J^{2/3}\frac{\partial^2 \bar{x}^j}{\partial x^l\partial x^n}
\right].
\end{equation}

A time derivative of this term appears in the shift evolution
equation~(\ref{eq:shiftevol2}).  Instead of evaluating this term 
exactly, we found it convenient to replace this equation with
\begin{equation}
\partial_t B^i=\left(\frac{dr}{d\bar{r}}\right)^2 \partial_t
\bar{\tilde{\Gamma}}^i -\eta B^i,
\label{eq:shiftevolfish}
\end{equation}
since calculating the time derivative of the quantity
$\tilde{\Gamma}^i$ requires taking a complicated spatially-varying linear combination of
time derivatives of the spatial metric. Within the different fisheye regions
(i.e.~outside the transition zones), our expression reproduces the
``non-shifting-shift'' condition, and it is equivalent to
Eq.~(\ref{eq:shiftevol2}) in the area surrounding the BH itself.
Elsewhere this modification affects only the coordinates, and not
any physical quantities.

Due to the asymptotic behavior of various quantities in fisheye
coordinates, we must modify our boundary conditions in some cases in
order to reproduce the desired behavior in physical coordinates.  In
all cases where outgoing wavelike boundary conditions are used, we
evaluate all radii in physical coordinates, not fisheye coordinates.
The conformal factor $\phi$, which does not asymptotically approach
unity in fisheye coordinates, is converted into the
physical coordinate expression using Eq.~(\ref{eq:phibar}), at which
point boundary conditions are applied and the expression is converted
back into fisheye.

\section{The relativistic Bondi solution}
\label{app:bondi}

A thorough derivation of the exact analytic relativistic 
Bondi solution may be found in Appendix~G
of \cite{ShapTeuk}.  Here we briefly review some of the basic features and
most relevant equations.  We assume that a BH of mass $M$ is placed
within an infinite cloud of gas that has a rest-mass density $\rho_\infty$ and
fluid 3-velocity $v^i = 0$ at spatial infinity, $r \rightarrow
\infty$.  We take the gas to be adiabatic with adiabatic index $\Gamma$.
We can then solve the equations of relativistic hydrodynamics to find
the stationary spherical accretion flow onto the black hole.

\subsection{Review of key equations}
\label{app:bondi_equations}

For convenience, we will derive the equations determining the flow in
Schwarzschild coordinates, and then convert these to the isotropic
coordinates used throughout this paper.  We denote Schwarzschild radii
$\hat{r}$, the 4-velocity $\hat{u}^\alpha$, and the inwardly directed
radial component of the 4-velocity $\hat{u}\equiv-\hat{u}^r$.  In
terms of these we can recast the relativistic continuity and Euler
equations in conserved form 
\begin{eqnarray}
4\pi\rho_0\hat{u}\hat{r}^2&\equiv&\dot{M}={\rm const.}\label{eq:bondicon}\\
h^2\left(1-\frac{2M}{\hat{r}}+\hat{u}^2\right)&\equiv&h_\infty^2={\rm const.}\label{eq:bondieuler}
\end{eqnarray}
Here we define the specific enthalpy $h\equiv 1+\epsilon+P/\rho_0$
where $\epsilon$ is the internal energy of the fluid.  We will assume
a gamma-law equation of state  
\begin{equation} \label{eq:gammalaw}
P=(\Gamma-1)\rho_0\epsilon,
\end{equation} 
for which the enthalpy is $h=1+\Gamma\epsilon$.  Inserting the latter
into (\ref{eq:gammalaw}) yields
\begin{equation} \label{eq:gammalaw2}
P =(\Gamma-1)\rho_0 \, \frac{h-1}{\Gamma}.
\end{equation}
For adiabatic flow the gamma-law equation of state
(\ref{eq:gammalaw}) implies the polytropic relation $P = \kappa
\rho_0^{\Gamma}$, where $\kappa$ is a constant.  Combining this with
(\ref{eq:gammalaw2}) yields
\begin{equation}
P = \kappa\rho_0^{\Gamma}=(\Gamma-1)\rho_0\left(\frac{h-1}{\Gamma}\right).
\label{eq:polyeos}
\end{equation}
For the gamma-law equation of state (EOS) (\ref{eq:gammalaw}) the speed of
sound is given by
\begin{equation}
a \equiv \frac{1}{h^{1/2}} \left( \frac{dP}{d\rho_0} \right)^{1/2} 
= \left( \frac{\Gamma P}{\rho_0 h} \right)^{1/2}.  
\end{equation}
Combining the above expressions we then find the following relations between
the enthalpy and the speed of sound
\begin{eqnarray}
a^2&=& (\Gamma-1)\frac{h-1}{h},\\
h&=&1+\frac{a^2}{\Gamma-1-a^2}.\label{eq:hs}
\end{eqnarray}

All smooth solutions to the conservation laws (\ref{eq:bondicon}) and
(\ref{eq:bondieuler}) satisfying the EOS (\ref{eq:polyeos}) must
pass through a sonic point, since the flow is subsonic at large radii
but must be supersonic at the horizon.  It can be shown that
at the sonic point the radial velocity $\hat{u}_s$ must satisfy
\begin{equation}
\hat{u}_s^2=\frac{M}{2\hat{r}_s},\label{eq:us}
\end{equation}
and that the speed of sound at the sonic point is
\begin{equation}
a_s^2=\frac{\hat{u}_s^2}{1-3\hat{u}_s^2}.\label{eq:as}
\end{equation}
The accretion rate for the transonic solution is given uniquely by
\begin{equation}
\dot{M}=4\pi\rho_s\hat{u}_s\hat{r}_s^2=4\pi\lambda_s M^2\rho_{\infty}a_{\infty}^{-3},
\end{equation}
where $\rho_{\infty}$ and $a_{\infty}$ are the asymptotic density and
sound speed, respectively, and $\lambda_s=\lambda_s(\Gamma)$ is tabulated in
Table~14.1 of \cite{ShapTeuk} for values $1\le\Gamma\le 5/3$.  

\begin{figure}[ht!]
 \centering \leavevmode \epsfxsize=3in \epsfbox{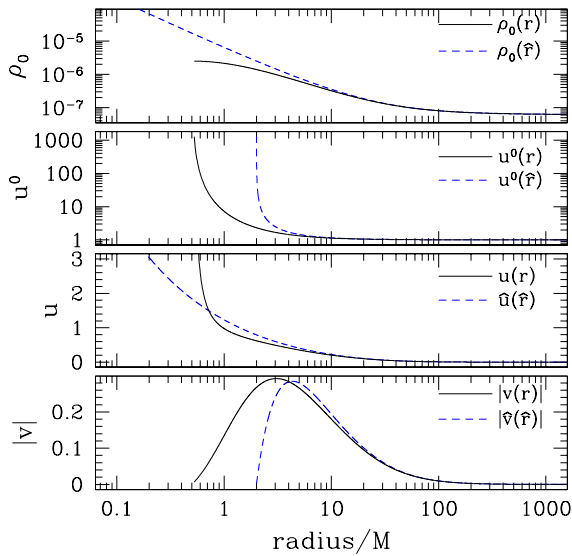}
 \caption{The Bondi solution for matter accreting onto a BH,
 where we set the fluid EOS to be a $\Gamma=4/3$ polytrope, the sonic
 radius as $\hat{r}_s=10M$, and the mass accretion rate
 $\dot{M}=10^{-4}$.  In the top two panels, we show the rest-mass density
 and $u^0$.  Solid curves show the quantity as a function of
 isotropic radii, whereas dashed show Schwarzschild radii, with the
 transformation given by Eq.~(\protect\ref{eq:isoschwarz}).  In the
 third panel, we show the radial component of the 4-velocity in both
 coordinate systems, with the transformation law given by
 Eq.~(\protect\ref{eq:uisoschwarz}).
In the bottom panel, we show the radial component of the 3-velocities
 for both coordinate systems.  We note that $u^0\rightarrow\infty$ and
 $v^i\rightarrow 0$ in either coordinate system as we approach the BH
 horizon, located at $r_h=0.5M$ in isotropic coordinates and
 $\hat{r}_h=2M$ is Schwarzschild.}
\label{fig:bondisol}
\end{figure}

In Fig.~\ref{fig:bondisol}, we show the particular Bondi solution we
use throughout this paper, corresponding to a $\Gamma=4/3$ EOS with a transonic
flow satisfying $\dot{M}=10^{-4}$ and $\hat{r}_s=10M$ (for
convenience, we set $M=1$).  In terms of these parameters, the
polytropic constant is given by $\kappa=7.56$, and the asymptotic
rest-mass density by $\rho_{\infty}=6.2\times 10^{-8}$.  In the top panel, we show 
the rest-mass density $\rho_0$ as a function of both
Schwarzschild radius $\hat{r}$ (dashed curve) and isotropic radius $r$
(solid curve).  Note that the isotropic solution terminates at
$r_h=0.5M$, since the interior Schwarzschild solution is mapped
through the throat of the BH onto the other sheet of the topology.  In
the second panel, we show the value of $u^0$ as a function of the two
coordinate radii, showing the divergence at the horizon.  In the third
panel, we show the radial component of the respective 4-velocities,
$u(r)$ and $\hat{u}(\hat{r})$.  Finally, in the bottom figure, we
show the radial component of the 3-velocities, $v\equiv |v^r|=|u^r|/u^0$, 
seeing that in both
cases this quantity goes to zero at the horizon, since the lapse
vanishes there.  We note, for clarity, that the rest-mass density
$\rho_0$ and $u$ remain
finite and smooth through the horizon in the Schwarzschild metric, 
becoming singular only at the
physical singularity at the origin. On the other hand, because of the
coordinate singularities present in the Schwarzschild metric, the
time-component of the 4-velocity $u^0$ and the radial 3-velocity $v^r$
both diverge at the horizon.

\subsection{Transformation to isotropic coordinates}
\label{app:bondi_iso}

To convert the Bondi solution from Schwarzschild to isotropic coordinates,
we only need to transform the radii and velocities.  The rest-mass
density -- as a time-component of the density 4-vector -- is
invariant under purely spatial coordinate transformations.  The radial
transformation is given by
\begin{eqnarray}
\hat{r}&=&r\left(1+\frac{m}{2r}\right)^2,\\
r&=&\frac{\hat{r}-M+\sqrt{\hat{r}(\hat{r}-2M)}}{2}.\label{eq:isoschwarz}
\end{eqnarray}
Since the flow is purely radial, we have
\begin{eqnarray}
u & = & -u^r = -\hat{u}^{\hat{r}}
\frac{dr}{d\hat{r}}
= \hat{u}\left(\frac{d\hat{r}}{dr}\right)^{-1} \nonumber \\
& = & \frac{\hat{u}}{(1-M/2r)(1+M/2r)}. \label{eq:uisoschwarz} 
\end{eqnarray}  
The 4-velocity's time-component $u^0$ remains unchanged under the
transformation, which means that the radial component of the
3-velocity $v^r = u^r/u^0$ transforms in the same way as the that
of the 4-velocity.  We find the 3-velocity from the normalization
\begin{equation}
-\alpha^2 (u^0)^2+\psi^4 (u^0)^2 v^2=-1.
\end{equation}

\subsection{Moving Bondi solutions}
\label{app:bondi_boost}
One approach to construct a moving Bondi solution is to view the
solution of the previous section from a frame in which the BH is
moving and described by a moving BH puncture spacetime.  Our strategy
is to adopt a moving puncture spacetime and regular matter density and
velocity profiles, with correct Bondi flow outer boundary conditions.
We then allow the solution to come to steady-state and use invariant
flow variables to compare with the stationary Bondi flow solution.

Specifically, we take the initial metric coefficients to be the vacuum
moving puncture solution, as described in Sec.~\ref{sec:numerical.ID}.
We take the initial density at any coordinate point to be
approximately the stationary isotropic Bondi solution.
Finally, we compute the initial velocity field using the special-relativistic
transformation law for velocities.  Denoting the stationary Bondi solution 
with $V$, and assuming a ``boost'' velocity  $v_b\equiv v_b^x$ 
in the $x$ direction, we have 
\begin{eqnarray}
\frac{v^x}{c}&=&\frac{v^x_b+V^x}{1+c^{-2}v^x_b V^x},\\
\frac{v^y}{c}&=&\frac{V^y}{\gamma_b(1+c^{-2}v^x_b V^x)},\\
\frac{v^z}{c}&=&\frac{V^z}{\gamma_b(1+c^{-2}v^x_b V^x)},
\end{eqnarray}
where
$\gamma_b\equiv (1-v_b^2/c^2)^{-1/2}$ and the local value of the speed of
light is $c\equiv \alpha/\psi^2$. Here, $v_b=P/{\mathcal M}$, where
$P$ is the momentum of the puncture.

The resulting initial data matches the Bondi solution only
approximately.  However, all deviations propagate away quickly,
leaving behind stationary Bondi flow onto a BH.

\end{appendix}

\bibliography{paper3bib}
\end{document}